# Large Language Model Agents for Evidence Based Genetic Disease Severity Classification

Tohid Ghasemnejad[1], Ahmadreza Argha[2], Mark Grosser[3], John Wang[4], Min Yang[5], Thantrira Porntaveetus[6], Tony Roscioli[7,8], Nigel H. Lovell[2], Mahmoud Aarabi*[9,10], Hamid Alinejad-Rokny*[1,11]

[1] UNSW BioMedical Machine Learning Lab (BML), School of Biomedical Engineering, UNSW Sydney, Sydney, NSW 2052, Australia

[2] School of Biomedical Engineering, UNSW Sydney, NSW 2052, Australia

[3] 23Strands, Pyrmont, Australia

[4] Division of Genetic and Genomic Medicine, Department of Pediatrics, University of Pittsburgh School of Medicine, Pittsburgh, PA, USA

[5] Shenzhen Institute of Advanced Technology, Shenzhen, China

[6] Center of Excellence in Precision Medicine and Digital Health, Department of Physiology, Faculty of Dentistry, Chulalongkorn University, Bangkok, Thailand

[7] New South Wales Health Pathology Genomics, Prince of Wales Hospital, Randwick, NSW, Australia

[8] Neuroscience Research Australia (NeuRA), University of New South Wales Sydney, NSW, Australia

[9] Medical Genetics and Genomics Laboratories, University of Pittsburgh Medical Center, Pittsburgh, PA, USA

[10] Departments of Pathology, and Obstetrics, Gynecology, and Reproductive Sciences, University of Pittsburgh School of Medicine, Pittsburgh, PA, USA

[11] Visiting Scholar (Collaborative Projects), Center of Excellence in Precision Medicine and Digital Health, Chulalongkorn University, Bangkok, Thailand

Correspondence to: H.A.R (h.alinejad@unsw.edu.au) and M.A. (mahmoud.aarabi@pitt.edu)

## Abstract

Classifying the severity of genetic conditions is subjective and slow, and screening panels vary widely in gene content. We built an artificial intelligence agent that retrieves PubMed abstracts, applies the severity tiers and quality-of-life criteria of professional screening guidelines, and cites a source for every claim, which a separate step checks. It classified 10,652 Human Phenotype Ontology terms. On 898 expert-labelled phenotypes, the agent assigned the correct category to 90.4% (95% CI 88.4–92.3%), against 73.2% for the same model without retrieved literature. Across tiers, 83.2% to 91.5% of its claims rested on direct evidence or a valid inference. Of 4,531 genes, 79.0% had a Severe or Profound condition, against 90.2% of genes on two carrier screening lists and 92.9% of genes linked to fetal or neonatal death. With retrieved literature, an AI agent can apply severity frameworks in close agreement with experts, and its cited evidence gives a transparent basis for choosing screening genes.

## Introduction

Artificial intelligence (AI) and other computational tools are now widely used in clinical genetics, especially to classify sequence variants. Following the establishment of the American College of Medical Genetics and Genomics and the Association for Molecular Pathology (ACMG/AMP) variant interpretation guidelines[1], automated tools have been developed to implement these standards, ranging from rule-based pipelines to AI-powered platforms and phenotype-driven prioritization systems[2–5]. Benchmarking studies have shown that these tools can identify disease-causing variants accurately, especially when they combine phenotype data with genomic features[6–8]. Despite this progress in classifying variants and interpreting genetic test results, several other steps in patient care are still done by hand. One example is the selection of genes for reproductive screening panels that are commonly used, prenatally or preconception, to determine the risk of couples for having a child affected by recessive genetic conditions[9]. Professional societies such as the American College of Medical Genetics and Genomics (ACMG) have set several requirements for including genes in carrier screening[10]. One is a carrier frequency of at least 1 in 200 for disease-causing (pathogenic or likely pathogenic) variants in population databases such as the Genome Aggregation Database (gnomAD). To address this requirement, lists of genes with a high carrier frequency have been generated from different gnomAD versions using well-established bioinformatic pipelines[11–14]. Another is that the recessive condition has at least moderate severity (moderate, severe or profound) in affected individuals, enough to justify screening heterozygous carriers who are at risk of having an affected child[10]. The American College of Obstetricians and Gynecologists (ACOG) also requires that the condition has a detrimental effect on quality of life (QoL)[15]. However, this requirement complicates the panel design process because determining disease severity is subjective and relies on varying expertise. Consequently, commercially available panels differ widely in size and gene content, and both carrier and newborn screening need a more objective, scalable way to assess severity[16,17].

Lazarin et al. developed a systematic framework for classifying disease severity[18]. From surveys of 192 healthcare professionals, they identified key disease characteristics and grouped them into four tiers by perceived impact, from Tier 1, the most severe (e.g., shortened life span in infancy), to Tier 4, the least severe (e.g., reduced fertility). The ACMG also uses this framework to classify genetic conditions by their characteristic profiles[10]. The authors noted, however, that manual classification is labor-intensive and may be biased for certain diseases[18]. To make ACOG's "detrimental effect on quality of life" criterion measurable, the Pediatric Quality of Life Inventory™ (PedsQL™) has been proposed[15,19]. PedsQL is an instrument for health-related quality of life that scores the functional impact of a health condition in four core domains (Physical, Emotional, Social and School/Cognitive Functioning), so a clinical description can be judged by its effect on daily life[20,21]. These problems grow with the number of phenotypes to be classified. The Human Phenotype Ontology (HPO) provides a standardized vocabulary for thousands of phenotypic abnormalities seen in human disease[22], and classifying the severity of each one by hand is slow and prone to inter-rater variability and subjective interpretation[19].

To reduce manual curation while keeping these established frameworks, we developed an AI agent that applies both the ACMG-endorsed severity tiers and the ACOG-compliant PedsQL™ functional impact model. The agent uses the ReAct (Reasoning and Acting) framework to reason step by step and Retrieval-Augmented Generation (RAG) to base each conclusion on retrieved PubMed literature (Figure 1). This lets it classify all 10,652 HPO terms that have a gene association. A recent study used GPT-4 to annotate every phenotypic abnormality in the HPO for nine clinical characteristics, each at four frequency levels, and benchmarked the output against labels derived from the HPO hierarchy[23]. Our approach differs in four ways: each classification is grounded in retrieved literature with checked citations, it applies the Lazarin tiers together with the PedsQL domains, it is validated against expert labels, and phenotypes are aggregated to gene–disease pairs that we compare with clinical gene lists. We aimed to make severity assessment of genetic conditions more objective and transparent, and to help clinical and laboratory experts reach consistent, evidence-based decisions. Here we show that the agent assigns the correct category to 90.4% of 898 expert-labelled terms, against 73.2% without retrieved literature, and that its gene-level classes rate genes on carrier screening lists as more severe than genes overall.

## Results

### Model Selection and System Performance

On 14 benchmark HPO terms, GPT-4o at temperature 0.2 and Claude 3.5 Sonnet had the highest accuracy (92.9%, 13 of 14 terms). GPT-4o was selected as the core model because it was faster (7.6 s against about 14 s per term) (Figure S1).

The validation set held 898 expert-labelled HPO terms whose category distribution matched that of all classified terms (PSI = 0.0303; Figure S3a). The agent assigned the correct category to 90.4% (95% CI 88.4–92.3%), with weighted precision of 91.0%, weighted recall of 90.4% and weighted F1 of 0.904 (Figure 2a). Agreement stayed high under class imbalance: MCC 0.889, Cohen's kappa 0.888, balanced accuracy 92.9% and macro-averaged F1 0.910. The most common errors were dysmorphic features classified as internal physical malformations (20 terms) or impaired mobility (16 terms) (Figure S3b). Per-category precision ranged from 73% to 100% and recall from 67% to 100%, with the lowest values for NFC (Figure 2b). For the ordinal tier, with Clinical Sign or Laboratory Abnormality, Out of Scope and NFC scored as one non-tiered class, accuracy was 91.2% (89.2–92.9%), quadratic-weighted kappa 0.926 (0.900–0.947), Spearman correlation 0.918, Kendall's tau 0.897 and mean absolute error 0.129 tiers; 97.1% of predictions were within one tier (Figure 2c). Per-tier F1 ranged from 0.867 (Tier 4) to 0.970 (non-tiered) (Figure 2d), and category accuracy within tiers from 86.6% (Tier 3) to 100% (Tier 1).

To measure what the retrieved literature adds, we classified the same 898 terms with the same model and prompt but without PubMed evidence. This baseline reached 73.2% category accuracy (70.2–76.1%) and 72.2% tier accuracy (69.0–75.1%; quadratic-weighted kappa 0.678). The agent's category accuracy was 17.3 percentage points higher than the baseline's (95% CI 14.5–20.0; McNemar test, $P = 1.7 \times 10^{-30}$), and its tier accuracy was 19.0 points higher (95% CI 16.0–21.8; $P = 1.2 \times 10^{-37}$; Figure 2e, f).

### Severity Tiers and Categories

Of the 10,652 classified HPO phenotypes, 2,966 (27.8%) were Tier 2, 2,351 (22.1%) Tier 3, 1,517 (14.2%) Tier 4 and 272 (2.6%) Tier 1, the most severe tier. The remaining 3,546 (33.3%) were non-tiered: 2,414 Clinical Sign or Laboratory Abnormality terms (for example, HP:0002315 Headache, HP:0004396 Poor appetite and HP:0003236 Elevated circulating creatine kinase concentration), 796 Out of Scope: Acquired/Multifactorial terms (for example, HP:0033235 Difficulty descending stairs, HP:0031446 Erosion of oral mucosa and HP:0032448 Achlorhydria) and 331 Not Further Classified (NFC) terms (for example, HP:0033454 Tube feeding, HP:0000118 Phenotypic abnormality and HP:0001470 Sex-limited expression); five further terms had no category recorded and are counted as NFC. The most common categories were internal physical malformations (n = 2,256) and dysmorphic features (n = 2,110) (category definitions in Supplementary Table 1). Tier distribution differed between categories (Figure 3a). Dysmorphic features were split between Tier 3 (41.6%) and Tier 4 (58.4%), whereas each shortened life span category fell in a single tier: infancy (n = 129) and childhood/adolescence (n = 50) in Tier 1, and adulthood (n = 61) in Tier 2.

### Evidence Quality and Validation

Most claims were backed by the literature: direct evidence and valid inferences together covered 83.2% to 91.5% of claims, depending on the tier (Figure 4a). Tier 1 had the highest direct-evidence rate (74.0%) and the lowest unsupported rate (5.7%), and Tier 4 the highest unsupported rate (14.1%; direct evidence 72.9%). Tiers 2 and 3 had direct-evidence rates of 73.6% and 71.7% and unsupported rates of 6.7% and 7.2%. Across categories, direct evidence ranged from 66.6% (hearing impairment) to 79.7% (intellectual disability) and exceeded 70% in 9 of the 13 categories; unsupported claims stayed below 10% in all 13 (Figure 4b).

Statement support scores clustered near the maximum in every tier and category (Figure 4c, d). Median scores were 0.97 for Tiers 1 and 2, 0.96 for Tier 3 and 0.95 for Tier 4, and Tier 4 had the longest tail of low scores (17.1% below 0.5). Intellectual disability had the tightest distribution (interquartile range 0.07) and dysmorphic features the widest (0.31). Support scores differed across tiers (Kruskal–Wallis $H(3) = 39.88$, $P = 1.1 \times 10^{-8}$) and across categories ($H(12) = 30.93$, $P = 0.0020$).

### Quality of Life Impact Assessment

PedsQL-based analysis rated 6,593 of 10,652 phenotypes (61.9%) as affecting quality of life, including 963 Clinical Sign or Laboratory Abnormality terms; among the 7,106 tiered phenotypes the figure was 5,630 (79.2%). Domain profiles differed across tiers ($\chi^2(12) = 1{,}020.84$, $P = 6.2 \times 10^{-211}$; Figure 5a; Table 1). Tier 1 had the most varied profile (Physical Functioning 47.7%, School/Cognitive Functioning 31.0%, Physical Symptoms 13.0%, Emotional Functioning 4.2%, Social Functioning 4.2%). Tier 2 was dominated by Physical Functioning (71.9%), with Physical Symptoms second (15.1%). Tier 3 was led by Physical Functioning (49.8%) but spread across School/Cognitive Functioning (19.3%), Social Functioning (11.5%), Emotional Functioning (10.2%) and Physical Symptoms (9.2%). Tier 4 was led by Emotional Functioning (38.3%), Physical Functioning (30.9%) and Social Functioning (23.5%), because all of its QoL-affected phenotypes are infertility terms (n = 44).

Category profiles also differed ($\chi^2(48) = 3{,}621.84$, $P = 1.1 \times 10^{-734}$; Figure 5b). Physical Functioning dominated internal physical malformations (65.1%), dysmorphic features (63.0%) and touch impairment (80.0%), and School/Cognitive Functioning dominated intellectual disability (57.9%). Shortened life span in infancy combined Physical Functioning (62.9%) with Physical Symptoms (22.6%), and in childhood/adolescence with School/Cognitive Functioning (21.8%; Physical Functioning 56.3%). Mental illness combined School/Cognitive Functioning (31.6%) and Emotional Functioning (31.4%). Vision impairment split between Physical Functioning (44.9%) and School/Cognitive Functioning (40.6%), and hearing impairment between School/Cognitive Functioning (44.3%) and Social Functioning (42.7%).

### Interventions and Early Onset by Tier and Category

The number of intervention types rose with the number of QoL domains affected (Figure 6a): the mean was 0.65 for phenotypes with no domain listed (n = 4,198), 1.52 with one domain, 1.72 with two, 1.72 with three, 1.95 with four and 2.67 with all five (n = 3); a linear fit to these six means gave $R^2 = 0.868$ (P =

0.0068). Physical Functioning was the most frequently affected domain (n = 5,411), followed by Physical Symptoms (n = 1,349), School/Cognitive Functioning (n = 1,312), Emotional Functioning (n = 606) and Social Functioning (n = 575) (Figure 6b). Phenotypes rated QoL-affected had more intervention types than those rated not affected (mean 1.53 against 0.72; Mann–Whitney U test, $P = 2.7 \times 10^{-413}$); this comparison uses QoL status rather than the domain count, so its 0.72 differs from the 0.65 above.

By tier, Tier 1 had the most intervention types (mean 1.70 ± 0.93 SD), followed by Tier 3 (1.53 ± 0.82) and Tier 2 (1.45 ± 0.92), with Tier 4 lowest (0.46 ± 0.72) (Kruskal–Wallis $H(3) = 1{,}494.29$, $P = 1.0 \times 10^{-323}$; Figure 6c). The shortened life span categories had the most (1.88 to 2.00; Figure 6d), followed by immunodeficiency/cancer (1.69) and the sensory impairments (1.43 to 1.66); dysmorphic features had the fewest (0.75) (Kruskal–Wallis $H(12) = 1{,}040.31$, $P = 4.0 \times 10^{-215}$). Figure 3c, d give the mix of intervention types in each category and tier. Early onset was recorded most often for dysmorphic features, internal physical malformations and shortened life span in infancy, and least often for mental illness ($\chi^2(12) = 1{,}131.92$, $P = 7.9 \times 10^{-235}$; Figure 3b).

**Quality of Life Status and Management Profiles**

QoL impact differed across categories and tiers (Figure S2a, b; Supplementary Table 2). All phenotypes in the Tier 1 categories (intellectual disability and shortened life span in infancy and childhood/adolescence) and in two Tier 2 categories (impaired mobility and shortened life span in adulthood) were QoL-affected, as were 98.3% of internal physical malformations (2,218/2,256). The mixed-tier categories varied more: dysmorphic features 41.6% (878/2,110), immunodeficiency/cancer 84.4% (526/623), mental illness 93.6% (292/312), and vision, hearing and touch impairment 94.2% (524/556), 90.1% (64/71) and 98.5% (64/65). Infertility, a Tier 4-only category, was 47.3% (44/93). QoL status was associated with category ($\chi^2(12) = 2{,}785.03$, $P = 7.6 \times 10^{-592}$).

Management also differed between QoL-affected and unaffected phenotypes ($\chi^2(7) = 2{,}312.92$, $P = 7.8 \times 10^{-496}$; Figure S2c). For affected phenotypes, the most common intervention types were surgery (30.6%), pharmacological treatment (24.2%), monitoring (21.5%) and therapeutic support (16.7%), and no treatment was rare (1.7%). For unaffected phenotypes, no treatment (24.3%), monitoring (24.0%) and pharmacological treatment (23.1%) were about equally common.

**Operational Efficiency**

Statement support scores across the 10,651 scored classifications averaged 0.82 (SD = 0.27), with a median of 0.96; 34.9% of phenotypes scored the maximum of 1.0 (Figure S3c). The agent took a mean of 6.63 steps (SD = 1.60) per classification; 78% of terms took 6 steps and 17% took 10. Processing cost averaged $0.101 USD per phenotype (range $0.036–$0.197; SD = $0.018), with the middle 50% of classifications costing $0.089–$0.112 (Figure S3d). Cost rose with the number of agent steps (Pearson $r = 0.40$, $P = 2.5 \times 10^{-410}$). The verifier flagged 1,169 of the 10,652 classifications (11.0%) for review, and the remaining 9,483 (89.0%) passed.

**Gene-Level Severity Classification and Evaluation**

The gene–disease table held 8,467 pairs (5,015 genes, 7,639 diseases): 6,458 from MONDO (after removing 7 with disputed or refuted GenCC evidence) and 2,009 added from GenCC (after removing 1,151 GenCC-added pairs on 24 umbrella disease terms linked to more than 10 genes). We excluded 359 pairs supported only by a Limited GenCC classification. Of the remaining 8,108 pairs, 1,399 could not be classified (1,376 had no phenotype annotation, 17 only Out of Scope or NFC phenotypes and 6 no phenotype left after the frequency filter), leaving 6,709 pairs (4,531 genes, 6,281 diseases) with a severity class: 2,436 (36.3%) Profound, 2,518 (37.5%) Severe, 1,433 (21.4%) Moderate and 322 (4.8%) Mild (Figure 7a; Table 2). In total, 4,954 pairs (73.8%) were Severe or Profound.

Severity varied across 23 body systems (Figure 7b; Table 2). The highest Profound share was in abnormalities of prenatal development or birth (69.5% of 620 pairs), growth abnormalities (57.2% of 2,223) and the nervous system (55.1% of 4,247). The musculoskeletal and nervous systems had the most pairs (4,249 and 4,247). The lowest Profound shares were in the thoracic cavity (21.4% of 14), neoplasms (23.5% of 502) and blood and blood-forming tissues (29.4% of 1,226), although most pairs in these systems were still Severe or Profound.

Severity also differed by mode of inheritance (Figure 7c; 38 pairs without an inheritance annotation were left out). X-linked pairs had the highest Profound share (49.1%, 172 of 350), followed by autosomal recessive (42.6%, 1,493 of 3,507), autosomal dominant (28.1%, 711 of 2,531) and complex inheritance (19.4%, 55 of 283). Of the 3,507 autosomal recessive pairs, 2,767 (78.9%) were Severe or Profound, as were 286 of the 350 X-linked pairs (81.7%).

Of all 4,531 genes in the analysis, 79.0% had a Severe or Profound condition (Figure 7d). The share was higher for genes on carrier screening lists: 90.2% for both the ACMG carrier screening list and Mackenzie's Mission (112 and 1,242 genes with a matched condition). Human Intolerome genes with a fetal effect (736 genes) had the highest share, 92.9%. Among Human Intolerome genes, those lethal only before birth scored slightly lower than those lethal only after birth. Of the 177 genes lethal only before birth, 89.3% had a Severe or Profound condition and 58.8% a Profound one; of the 395 genes lethal only after birth, the figures were 93.4% and 64.1%. ACMG secondary findings genes (81) were below the base rate, at 74.1%; this list is selected for medical actionability rather than for severity, and its conditions are mostly cancer predisposition and cardiovascular disorders. A phenotype in the category shortened life span in infancy or childhood/adolescence was present for 26.5% of all genes, 34.1% of Mackenzie's Mission genes, 42.9% of ACMG carrier screening genes, 43.2% of ACMG secondary findings genes and 52.2% of Human Intolerome genes (Figure 7e).

## Discussion

We developed and validated an LLM agent that classifies the severity of genetic conditions from HPO records and the published literature, and applied it to all 10,652 HPO phenotype terms with a gene association. Each term received a severity tier, a quality-of-life profile, an age of onset and a management profile, and the classes agreed with expert labels for most phenotypes.

Screening panels vary widely. Surveys of carrier screening panels show little agreement on gene content. Among 16 providers in 2017, the number of conditions screened ranged from 41 to 1,792, only three conditions were screened by all of them, and only about 10% of genes were included by at least half of the providers[16]. A 2023 review of 22 panels found the same pattern: only 15 of 2,205 genes (0.7%) were on every panel, and 695 (31.5%) were on just one[24]. In recent years, the ACMG recommendation has become the most widely adopted approach, which selects genes by carrier frequency and severity[10,25]. Carrier frequency has been shown to be a dynamic metric that is subject to change based on the degree of consanguinity, founder effects, bottlenecks, and the availability of genome data from new larger populations[11,12,26]. Disease severity is less prone to change than carrier frequency as it originates from well-established clinical presentations in the literature. Many panels can also be narrowed to a smaller set of genes for a patient's clinical needs[24], and severity classes could help laboratories choose genes for both broad and targeted panels. The clinical utility of carrier screening using a carefully selected gene panel is demonstrated by the recent Mackenzie's Mission study, which identified that 1.9% of 9,107 couples had a newly identified increased risk for a severe genetic condition and 76.6% of them used or planned to use reproductive interventions[27,28]. This result supports screening with broad panels. Genomic newborn screening faces the same problem at a larger scale. Across 27 newborn sequencing programmes, the number of genes analysed ranged from 134 to 4,299, and only 74 genes (1.7%) were included by more than 80% of programmes[29]. Severity is one of the core criteria for these lists, together with treatability, penetrance and age of onset[17], so a reproducible severity classification could help programmes align their gene lists, and the management and onset profiles recorded for each phenotype could inform the treatability and onset criteria.

The agent's agreement with the expert labels depended on the literature it retrieved. Murphy et al. queried GPT-4 directly and reported recall of 89% to 100% against phenotypes that the HPO hierarchy places under each clinical characteristic[23], but some of these test terms share words with the characteristic itself. Our test asked more of the model: it had to choose one of 16 categories for each term, scored against expert labels, and without PubMed evidence the same model lost 17.3 percentage points of category accuracy. Recall against the ontology structure may therefore overstate how well an LLM applies a severity framework without evidence. Murphy et al. noted that their annotations left out treatment availability and the psychological and social burden of a phenotype[23]. Our framework records both, through the management profile and the PedsQL domains. The QoL-affected Tier 4 phenotypes show why this matters: all 44 are infertility terms, and emotional impact is their most common affected domain, which the tier alone does not show.

The Lazarin tiers reflect the views of healthcare professionals[18], and parents may weigh disease characteristics differently. Sajko et al. asked 320 expectant women and their partners to rate 28 disease characteristics, and they gave a shortened life span more weight than intellectual disability, although the Lazarin framework puts the two in the same tier[30]. Because each tier is set from the phenotype's category and QoL status by fixed rules, a screening programme could map the same categories onto a parent-derived ranking such as this one without classifying the terms again. The domain profiles may also help counsellors answer what parents want to know about their child's quality of life[31].

LLM citations often fail to support the claims they are attached to: across seven LLMs and 800 medical questions, Wu et al. found that 50% to 90% of answers were not fully supported by their cited sources, and even for GPT-4o with web search about 30% of individual statements were unsupported[32]. In our agent, 5.7% to 14.1% of claims were unsupported, depending on the tier. The settings differ, so the figures are not directly comparable, but giving the model a fixed set of retrieved sources and requiring a citation to one of them for every claim probably kept this rate down. The verification step flagged 11.0% of classifications, which gives curators a defined set to review first.

At the gene level, X-linked and autosomal recessive pairs were more often Profound than autosomal dominant pairs. Two earlier studies found that autosomal dominant genes are less often linked to early or severe disease. Among Mendelian genes in OMIM, only 12% of genes with records of prenatal, neonatal or infant death were autosomal dominant, against 34% of genes with no record of premature death[33]. Among genes for inborn errors of immunity, 76.5% of autosomal recessive and 75.0% of X-linked recessive genes caused early-onset, highly penetrant disease, while 80.9% of autosomal dominant genes caused later-onset or incompletely penetrant disease. The authors of that study note that this difference may partly reflect which dominant diseases are easier to discover[34]. Carrier screening looks for autosomal recessive and X-linked conditions, and only 5.1% of our autosomal recessive pairs and 2.6% of X-linked pairs were Mild (Figure 7c), the only class below the ACMG threshold of at least moderate severity[10]. On its own, that threshold would therefore exclude few autosomal recessive or X-linked conditions from carrier screening. For couples, the difference between Moderate and Severe matters. In a survey of 64 at-risk couples whose conditions were classed with the Lazarin algorithm, 76% of those carrying a Severe or Profound condition took or planned reproductive action, against 22% of those carrying a Moderate condition, whereas inclusion of the condition in screening guidelines made no difference[35]. Our classes separate Moderate from Severe, and conditions with a phenotype that shortens life in childhood from those without, and these distinctions may be more useful for counselling and tiered panels than a single threshold.

The phenotype-level classes have four limitations. First, the agent classifies structured HPO terms, which may not capture all the clinical detail in narrative descriptions, and some phenotypic features have no suitable HPO term. Second, each class reflects the literature retrieved when the term was classified, so terms will need to be reclassified as evidence accumulates. Third, 3,546 terms (33.3%) received no tier. Most were clinical signs or laboratory abnormalities (2,414) or acquired or multifactorial phenotypes (796), which the framework leaves untiered by design. The terms that fit no category (NFC), however, may point to gaps in the Lazarin categories. Fourth, Tier 4 had the highest share of unsupported claims (14.1%, against 5.7% to 7.2% in Tiers 1 to 3). This may be because most Tier 4 phenotypes (1,473 of 1,517) are placed there for having no functional impact, and the literature seldom describes the absence of an impact.

The gene-level classes have two limitations. First, a class depends on how fully its disease is annotated: the counting rules favour diseases with many annotated phenotypes, and the 30% frequency filter removes phenotypes only when a frequency is recorded. A sparsely annotated disease may therefore be rated less severe than an equally severe disease with rich annotation. Second, Human Intolerome genes lethal only before birth scored slightly lower than those lethal only after birth, probably because HPO annotations mostly describe disease after birth.

The next steps follow from the limitations of this study. Classifying the clinical descriptions in case reports, and not only HPO terms, would capture detail the ontology lacks, and prenatal phenotype annotations would let the gene-level classes reflect death before birth. Rerunning the agent with each HPO release would keep the classes current as the literature grows. Mapping its categories onto rankings made by parents would show how much the choice of framework changes which genes look severe. The study shows that an LLM agent given retrieved abstracts can apply the Lazarin tiers and PedsQL criteria to HPO terms: its categories matched the expert labels for most validation terms. Each claim cites one of the retrieved sources, and a separate step checks those citations, so a curator can see why a term received its class. We see the agent as a first pass: curators check the flagged terms, and clinicians make the final decisions.

## Methods

### Study Workflow and Severity Tiers

Figure 1 shows the workflow. It combines the hierarchical severity tiers used by the ACMG[10,18] with a QoL evaluation based on PedsQL™ domains, age of onset, and treatment availability, following ACOG recommendations[15]. These classification criteria were provided to a ReAct agent that executed five sequential phases through iterative reasoning loops: contextualisation, query generation, evidence retrieval, adaptive expansion and classification. Phenotype term classifications were then mapped to gene and disease associations and aggregated through a hierarchical algorithm to produce gene-level severity designations: Profound, Severe, Moderate, or Mild.

Adapting the severity framework developed by Lazarin et al.[18], we assigned each of the 10,652 HPO terms with a phenotype–gene association to one of four severity tiers. Tier 1, the most severe, covered shortened life span in infancy or childhood/adolescence and intellectual disability. Tier 2 covered shortened life span in adulthood, impaired mobility and internal physical malformations. Tier 3 covered sensory impairment, immunodeficiency/cancer, mental illness and dysmorphic features, and Tier 4 covered infertility. Phenotypes in Tier 1 and Tier 2 categories were always counted as affecting quality of life. A phenotype in a Tier 3 category stayed in Tier 3 if it affected quality of life and moved to Tier 4 if it did not, so the tier follows from the category and the QoL status by fixed rules. Full category definitions are in Supplementary Table 1.

Three non-tiered designations handled terms outside this framework. "Clinical Sign or Laboratory Abnormality" captured objective findings such as transient symptoms and laboratory results rather than persistent functional disability. "Out of Scope: Acquired/Multifactorial" covered phenotypes that are primarily acquired, such as hypertension. "Not Further Classified" (NFC) was applied to terms that did not fit any Lazarin category. Terms in these categories did not receive a numerical tier.

### Quality of Life Assessment, Management, and Age of Onset

ACOG does not define how to judge its criterion of a "detrimental effect on quality of life". Following Arjunan et al., who defined quality of life by the domains of the Pediatric Quality of Life Inventory (PedsQL) when interpreting this criterion[19], we used the domains of the PedsQL 4.0 Generic Core Scales and Infant Scales, a validated instrument for children and infants[20,21]. The agent mapped evidence to five domains: Physical Functioning, Emotional Functioning, Social Functioning, School/Cognitive Functioning and Physical Symptoms[21]. The last is a domain of the Infant Scales and covers gastrointestinal, respiratory and skin symptoms. Supplementary Table 3 lists the items of the PedsQL Infant Scales.

The QoL framework had two roles. First, it determined whether each phenotype met the ACOG QoL criterion. Second, for phenotypes in Tier 3 categories, it decided between Tier 3 and Tier 4. For example, orofacial cleft (HP:0000202) and thick eyebrow (HP:0000574) are both dysmorphic features, but orofacial cleft affects Physical Functioning (e.g., feeding difficulties)[36] and Social Functioning (e.g., speech impairment)[37], and is classified as QoL-affected (Tier 3), whereas thick eyebrow is an isolated cosmetic trait with no functional impact and is classified as not affected (Tier 4).

Two further ACOG criteria were recorded. For onset early in life, early onset was recorded when the agent described the onset as congenital, at birth, neonatal, in infancy, in childhood or paediatric. For the criterion that a condition requires surgical or medical intervention, the agent extracted management strategies from the literature into eight categories (Surgical, Pharmacological, Therapeutic Support, Monitoring, Dietary, Assistive Devices, Palliative Care and No Treatment), or stated that the sources gave no information on management. No Treatment was not counted as an intervention.

### Model Selection and Optimization

We benchmarked six LLMs (GPT-5, Grok-4, Gemini-2.5-pro, Claude-3.5-sonnet, GPT-4o, and GPT-4o-mini) at three temperatures (0.0, 0.2, 0.5) and three top-p values (0.90, 0.95, 1.0) on 14 expert-labelled HPO terms spanning the tiers and categories. A term was scored correct only when its tier, category and quality-of-life rating all matched the expert label, accuracy was averaged over the three top-p values, and processing time per term was recorded. All 14 terms are also among the 898 validation terms. GPT-4o was accessed through OpenRouter under the model identifier openai/gpt-4o and was used at temperature 0.2 for the agent, the verification agent and the baseline.

### Prompt Engineering

A structured prompt was designed to guide an LLM through a defined clinical reasoning process, aiming to classify an HPO term into predefined severity categories based on Lazarin et al.[18] and to assess its impact on QoL domains[20,21]. All inputs (HPO context and literature) were supplied in a standardised JSON structure with a unique identifier per source, which the LLM cited for every factual claim. The model first determined whether the phenotype was congenital or acquired; congenital phenotypes were then traversed through a hierarchical decision tree and assigned to the first applicable category, followed by a PedsQL-based QoL assessment. Mandatory citation and a prohibition on internal knowledge kept reasoning grounded in the supplied evidence. The output was a single JSON object containing the reasoning trace, classification, management profile, and cited references, so it could be parsed and checked automatically.

### Agent Workflow

Each HPO term was processed through five sequential phases. In Phase 1 (Term Contextualisation), the system retrieved the full definition, synonyms, and hierarchical relationships from the HPO database (release 2025-05-06) and mapped the term to its corresponding Medical Subject Heading (MeSH) via NCBI to improve search precision. In Phase 2 (Query Generation), the LLM expanded the vocabulary with common clinical diagnoses and synonyms, which were combined with HPO synonyms and the MeSH term. Five predefined search templates (congenital and genetic associations, hereditary and familial patterns, paediatric and neonatal onset, childhood prognosis, and management of inherited disorders) were then applied to produce up to 15 deduplicated queries per term. In Phase 3 (Evidence Retrieval), PubMed was searched hierarchically, first combining the MeSH term with genetic-disease keywords and progressively broadening to include all synonyms, then removing context filters, keeping at most 10 abstracts per search. In Phase 4 (Adaptive Expansion), triggered when fewer than five relevant articles were found, the LLM identified commonly associated genetic syndromes and repeated the search to find more evidence for rare conditions. In Phase 5 (Classification), the system synthesised retrieved evidence, applied the hierarchical rules, assigned a category based on the primary manifestation, and determined QoL status. The evidence for each term was its HPO record and the retrieved PubMed abstracts (mean 8.0 abstracts per term). Of the 10,652 terms, 10,211 were classified with HPO release 2025-05-06. The other 441 were added to the term set later and were classified on 20 November 2026 with HPO release 2026-09-01.

**Source Verification System**

A separate verification agent assessed the factual accuracy of each classification using a modified SourceCheckup framework[32]. In the extraction stage, claims were taken from the reasoning trace by rule: every sentence ending in a reference tag in the functional impact, severity and conclusion sections, with its tag. In the verification stage, the same model (GPT-4o, temperature 0.2), called separately as a fact-checker, matched each claim to its cited source and assigned a weighted support score: direct statement from the source (1.0), valid medical inference (0.90), weak inferential support (0.60), or absence of support (0.0). The support score of a term is the mean of these weights over its claims. A classification was flagged for review when its support score was below 0.5.

**Phenotype-Level Performance Evaluation**

We built a ground truth set of HPO terms, selected by hand, of phenotypes that covered every category in proportion to how often that category occurred among all classified terms. To check that the set was representative, we compared its category distribution with the agent's classifications of all classified terms using the Population Stability Index (PSI):

$$\mathrm{PSI} = \sum_{i=1}^{K} \left(p_{\mathrm{actual},i} - p_{\mathrm{expected},i}\right) \times \ln\left(\frac{p_{\mathrm{actual},i}}{p_{\mathrm{expected},i}}\right) \qquad \text{(Equation 1)}$$

where $K$ is the number of categories, $p_{\mathrm{actual},i}$ is the proportion of category $i$ in the ground truth set and $p_{\mathrm{expected},i}$ is the proportion of the agent's classifications in category $i$; proportions of zero were set to 0.0001. The target number of terms in each category was

$$n_{\mathrm{target},i} = \mathrm{round}\left(p_{\mathrm{expected},i} \times N\right) \qquad \text{(Equation 2)}$$

where N is the size of the ground truth set. In each category, terms were added by hand until its target count was reached. Labels were assigned in two steps. First, one curator (TG) proposed a category and tier for each term. Two clinical and laboratory geneticists (MA and JW) then each reviewed every proposed label and marked it as agree or disagree; disagreements were resolved by consensus. Separately, the geneticists checked the agent's reasoning for a subset of terms. Fifteen labelled terms were not among the 10,652 classified terms (11 mode-of-inheritance terms, the clinical modifier Chronic and three other phenotype terms) and were removed, leaving 898 terms.

We measured agreement with the expert labels for the category and for the tier, using accuracy and Cohen's kappa:

$$\kappa = \frac{p_o - p_e}{1 - p_e} \quad \text{(Equation 3)}$$

where $p_o$ is the observed agreement and $p_e$ is the agreement expected by chance. Kappa was unweighted for the category and quadratic-weighted for the tier, with weights $w_{ij} = (i-j)^2/(T-1)^2$, where $T$ is the number of tier classes, and $\kappa_w = 1 - \sum w_{ij} o_{ij} / \sum w_{ij} e_{ij}$, where $o_{ij}$ and $e_{ij}$ are the observed and chance-expected proportions of terms with true tier $i$ and predicted tier $j$. Tiers were scored in five ordered classes: Tier 1 to Tier 4, then one non-tiered class for Clinical Sign or Laboratory Abnormality, Out of Scope and NFC. Because the tier is derived from the category and the QoL status, tier agreement is not independent of category agreement: a category error changes the tier unless both categories share a tier. For the category we also report the Matthews correlation coefficient (MCC) in its multi-class form[38]:

$$\text{MCC} = \frac{c \times s - \sum_k p_k t_k}{\sqrt{(s^2 - \sum_k p_k^2)(s^2 - \sum_k t_k^2)}} \quad \text{(Equation 4)}$$

where $c$ is the number of terms classified correctly, $s$ is the number of terms, and $p_k$ and $t_k$ are the numbers of terms predicted to be in and truly in category $k$. We also report balanced accuracy, the mean recall over the categories:

$$\text{Balanced accuracy} = \frac{1}{K} \sum_{k=1}^{K} \frac{TP_k}{TP_k + FN_k} \quad \text{(Equation 5)}$$

and the macro-averaged and weighted F1 scores:

$$F1_{\text{macro}} = \frac{1}{K} \sum_{k=1}^{K} F\,1_k \quad \text{(Equation 6a)}$$

$$F1_{\text{weighted}} = \sum_{k=1}^{K} \frac{n_k}{s} F1_k \quad \text{(Equation 6b)}$$

where $TP_k$ and $FN_k$ are the true positives and false negatives for category $k$, $F1_k$ is its F1 score and $n_k$ is the number of terms whose expert label is k. For the tier we also report Spearman's correlation, Kendall's tau and the mean absolute error between true and predicted tiers.

To test what the retrieved literature adds, we classified the 898 terms again with the same model (temperature 0.2) and the agent's classification prompt, giving each term's name, definition, synonyms and comment from the HPO but no PubMed abstracts. The source list and the rules that refer only to sources were removed, and one instruction was added: "No sources are provided for this term. Wherever these instructions refer to sources or evidence, use your own clinical knowledge instead." The baseline was also asked to include a short definition of the phenotype in its answer. The tier was set by the same rules as for the agent, and the baseline was scored against the same expert labels.

### Gene-Level Severity Classification

We took germline gene–disease relationships from MONDO[39] (release 2026-09-01): "has material basis in germline mutation in" (RO:0004003) and "has material basis in gain of function germline mutation in" (RO:0004001), stated directly or inside the logical definition of an active disease term. We added pairs with at least one Definitive, Strong or Moderate classification in the Gene Curation Coalition (GenCC) database[40] (export of 15 November 2026) for the same HGNC and MONDO identifiers, unless the MONDO term was obsolete or absent in this release. GenCC-added pairs on umbrella disease terms linked to more than 10 genes were removed, because such terms pool the phenotypes of many genetic forms. We also removed pairs whose strongest GenCC classification was Disputed Evidence, Refuted Evidence or No Known Disease Relationship, and excluded pairs whose strongest classification was Limited.

We used the positive phenotype annotations (aspect P, no NOT qualifier) of HPO release 2026-09-01[22]. The OMIM or Orphanet disease of each annotation was mapped to MONDO through skos:exactMatch cross-references, only when it matched exactly one active MONDO term. HPO frequency terms were converted to numbers (Obligate 1.0, Very frequent 0.9, Frequent 0.55, Occasional 0.17, Very rare 0.025, Excluded 0); counts (n/m) and percentages were used as given, and each disease–phenotype pair took the highest frequency across its annotations. A phenotype whose every annotation gave a frequency of 0 was treated as absent and removed. We also removed phenotypes whose highest frequency was below 0.30, and kept phenotypes without frequency information.

To avoid counting one finding twice, a Tier 1, 2 or 3 phenotype was not counted when a more specific phenotype of the same pair (a descendant through is_a links, at any depth) had the same tier.

Each phenotype kept the tier assigned by the agent. For each gene–disease pair $(g, d)$ we counted the distinct phenotypes of each tier:

$$n_{\text{Tier } i} = |\{p \in P(g, d) : \text{tier}(p) = i\}|, \quad i \in \{1,2,3\} \qquad \text{(Equation 7)}$$

where $P(g, d)$ is the set of phenotypes kept for the pair after the steps above. Tier 4 and non-tiered phenotypes do not count toward the class, and pairs whose phenotypes were all NFC or Out of Scope were excluded. Each pair took the first class whose rule it met, checked in this order (Figure S4 gives a worked example):

- Profound: $n_{\text{Tier1}} > 1$.
- Severe: $n_{\text{Tier1}} = 1$, or $n_{\text{Tier2}} \geq 1$ and $n_{\text{Tier2}} + n_{\text{Tier3}} \geq 4$.
- Moderate: $n_{\text{Tier2}} \geq 1$ or $n_{\text{Tier3}} \geq 1$.
- Mild: no counted Tier 1, 2 or 3 phenotype (only Tier 4 or Clinical Sign phenotypes).

Each pair took its mode of inheritance from the HPO inheritance annotations of its disease or, when HPO had none, from GenCC submissions. Pairs were grouped as autosomal dominant, autosomal recessive, X-linked, or complex (more than one of these modes, or only other inheritance terms). For body systems, each phenotype was assigned to every top-level HPO category it falls under (the 23 children of Phenotypic abnormality, HP:0000118), and a pair was counted under every body system that held at least one of its phenotypes.

**Comparison with Clinical Gene Lists**

We compared the gene-level classes with four lists: the ACMG secondary findings list v3.2 (81 genes)[41], the ACMG Tier 3 carrier screening list (113 genes)[10], the Mackenzie's Mission list (1,300 genes)[27] and the Human Intolerome[42]. From the Human Intolerome we used genes whose lethality acts through the fetus (gene effect Fetal or Maternal,Fetal) with Known or Emerging human evidence (824 genes). None of these lists grades severity, so we tested whether listed genes were rated more severe than genes overall, not exact agreement. List genes were resolved to HGNC identifiers, and each listed condition's MIM number was mapped to MONDO through skos:exactMatch. A listed gene–condition pair was matched when the gene was paired with that MONDO term, with its replacement if the term was obsolete, or with a broader or narrower MONDO term of the same condition. As a base rate, every gene in the analysis took the highest class over all its pairs. As a second measure, we counted genes whose condition has at least one phenotype in the category shortened life span in infancy or in childhood/adolescence.

**Statistics and Reproducibility**

No statistical method was used to predetermine sample size. The analysis covers all 10,652 HPO terms annotated to diseases with a gene association in the HPO releases used in this study, and the ground truth terms were selected by hand to match the category distribution of all classified terms. Exclusions and their reasons are described in the sections above, with counts in the Results. Each term was classified once.

All tests were two-sided with a significance threshold of $\alpha = 0.05$. Kruskal–Wallis H tests compared continuous or ordinal outcomes across more than two groups (for example, management complexity across tiers or categories). A Mann–Whitney U test compared QoL-affected with unaffected phenotypes. Pearson's chi-square test of independence assessed associations between categorical variables. Phenotypes mapped to several QoL domains count once per domain, so the chi-square tests on domain counts (Figure 5) treat mappings rather than phenotypes as observations. The relationship between QoL domain breadth and mean intervention count was fitted by ordinary least squares regression and reported as $R^2$; the 95% confidence band in Figure 6a comes from 1,000 bootstrap resamples (seaborn, seed 42). Pearson's correlation measured the association between processing cost and the number of agent steps. Confidence intervals (95%) for accuracy, kappa and the agent–baseline differences came from 2,000 bootstrap resamples of terms (percentile method, seed 42), and the agent and baseline were compared on the same terms with an exact McNemar test. No correction for multiple testing was applied; across the 14 tests, a Bonferroni threshold would be $P < 0.0036$. Exact P values are reported to two significant figures, and those too small for standard floating-point arithmetic were computed on the log scale. Test statistics, degrees of freedom and n are given in the Results and figure legends. Error bars show standard deviation unless stated otherwise. Analyses were performed in Python 3.13 with NumPy 2.5, SciPy 1.18, scikit-learn 1.9 and pandas 3.0, and figures were made with Matplotlib 3.11 and seaborn 0.13.

## Data Availability

The Human Phenotype Ontology[22] (https://hpo.jax.org/; release 2025-05-06, read by the agent for 10,211 terms, and release 2026-09-01, read by the agent for the other 441 terms and used with its disease annotations for the gene-level analysis), MONDO[39] (release 2026-09-01), the GenCC submissions[40] (export of 15 November 2026) and the HGNC gene set[43] (downloaded 18 November 2026) are publicly available. PubMed abstracts and MeSH terms used for evidence retrieval were obtained via NCBI E-utilities. The reference panels used for external validation are available from the original publications: the ACMG carrier screening panel[10], the ACMG Secondary Findings v3.2 list[41], Mackenzie's Mission[27] and the Human Intolerome[42]. Copies of the ontology, GenCC, HGNC and gene-list files used in this study are included in the Zenodo record below; their reuse is subject to the terms of the original sources. The list of 10,652 HPO terms, the expert labels of the ground truth set, the model-selection results, the full per-term classification output for all 10,652 terms (category, tier, QoL domains, management, reasoning, citations and support scores), the no-retrieval baseline output and the gene–disease table with the severity class of each pair are deposited at Zenodo (https://doi.org/10.5281/zenodo.24463716) and at https://github.com/T0hid/severity-hpo-classification.

## Code Availability

The code is available at https://github.com/T0hid/severity-hpo-classification and archived at Zenodo (https://doi.org/10.5281/zenodo.24463716). The repository contains the classification agent and its verification agent, the no-retrieval baseline, the model-selection analysis, the gene–disease preprocessing and severity aggregation, the comparison with clinical gene lists and the figure scripts. It also includes the saved outputs of the run reported here, a command that recomputes the numbers in this manuscript from these outputs without calling a language model, and a Dockerfile for the software environment. New classification runs need an OpenRouter API key, which the code reads from an environment variable.

## Author Contributions

TG conceptualised the study, designed and developed the ReAct–RAG agent framework, implemented the classification pipeline, performed all computational analyses, and wrote the first draft of the manuscript. AA provided input on methodology and reviewed the manuscript. MG contributed to manuscript editing and revision. NHL contributed to the study design, provided oversight on the biomedical engineering aspects, and reviewed the manuscript. JW contributed to the clinical validation of the phenotype classifications and reviewed the manuscript. MY contributed to methodology review. TP contributed to clinical review. TR provided clinical genetics expertise and reviewed the manuscript. MA contributed to the conceptualisation of the clinical framework, provided expert curation of gene and phenotype datasets for validation, supervised the clinical genetics aspects of the study, and critically revised the manuscript. HAR conceptualised and supervised the study, contributed to the study design and analytical framework, and revised the manuscript.

## Acknowledgements

HAR was supported by the UNSW Scientia Program Fellowship and the Australian Research Council Discovery Early Career Researcher Award (DECRA) under grant DE220101210. HAR was also supported by ARC Future Fellowship under grant FT260100970. This collaborative project was also supported by the Second Century Fund (C2F), Chulalongkorn University. Additionally, the project has support from the Ratchadaphiseksomphot Endowment Fund, Top-up Grant of C2F, Chulalongkorn University. The authors used Claude (Anthropic) to edit the language of the manuscript and improve its clarity. The authors reviewed all edits and take full responsibility for the content.

## Competing Interests

UNSW Sydney holds intellectual property rights related to the ReAct–RAG autonomous agent framework and classification pipeline, the ACMG severity classification framework, the ACOG quality-of-life assessment module (PedsQL™ domain mapping and functional impact evaluation), the management and therapeutic strategies for each HPO term, the source verification system, and the gene-level severity aggregation algorithm described in this work. The ACMG-endorsed severity classification framework is licensed to 23Strands Pty Ltd. MG is CEO of 23Strands Pty Ltd. TG reports a top-up scholarship from 23Strands. All other authors declare no competing interests.

## Materials & Correspondence

Correspondence and requests for materials should be addressed to Hamid Alinejad-Rokny (h.alinejad@unsw.edu.au) and Mahmoud Aarabi (mahmoud.aarabi@pitt.edu).

**References**

1. Richards, S. *et al.* Standards and guidelines for the interpretation of sequence variants: a joint consensus recommendation of the American College of Medical Genetics and Genomics and the Association for Molecular Pathology. *Genet. Med.* **17**, 405–424 (2015).
2. Nicora, G., Zucca, S., Limongelli, I. & Bellazzi, R. A machine learning approach based on ACMG/AMP guidelines for genomic variant classification and prioritization. *Sci. Rep.* **12**, 2517 (2022).
3. Stawiński, P. & Płoski, R. Genebe.net: Implementation and validation of an automatic ACMG variant pathogenicity criteria assignment. *Clin. Genet.* **106**, 119–126 (2024).
4. Nicora, G. *et al.* CardioVAI: An automatic implementation of ACMG-AMP variant interpretation guidelines in the diagnosis of cardiovascular diseases. *Hum. Mutat.* **39**, 1835–1846 (2018).
5. Kopanos, C. *et al.* VarSome: the human genomic variant search engine. *Bioinformatics* **35**, 1978–1980 (2019).
6. Ghasemnejad, T. *et al.* Comprehensive evaluation of ACMG/AMP-based variant classification tools. *Bioinformatics* **42**, btaf623 (2026).
7. Kelly, C. *et al.* Phenotype-aware prioritisation of rare Mendelian disease variants. *Trends Genet.* **38**, 1271–1283 (2022).
8. Yuan, X. *et al.* Evaluation of phenotype-driven gene prioritization methods for Mendelian diseases. *Brief. Bioinform.* **23**, bbac019 (2022).
9. Kraft, S. A., Duenas, D., Wilfond, B. S. & Goddard, K. A. B. The evolving landscape of expanded carrier screening: challenges and opportunities. *Genet. Med.* **21**, 790–797 (2019).
10. Gregg, A. R. *et al.* Screening for autosomal recessive and X-linked conditions during pregnancy and preconception: a practice resource of the American College of Medical Genetics and Genomics (ACMG). *Genet. Med.* **23**, 1793–1806 (2021).
11. Schmitz, M. J. *et al.* Carrier frequency of autosomal recessive genetic conditions in diverse populations: Lessons learned from the genome aggregation database. *Clin. Genet.* **102**, 87–97 (2022).
12. Schmitz, M. J. *et al.* Leveraging diverse genomic data to guide equitable carrier screening: Insights from gnomAD v.4.1.0. *Am. J. Hum. Genet.* **112**, 181–195 (2025).
13. Guo, M. H. & Gregg, A. R. Estimating yields of prenatal carrier screening and implications for design of expanded carrier screening panels. *Genet. Med.* **21**, 1940–1947 (2019).
14. Hotakainen, R., Järvinen, T., Kettunen, K., Anttonen, A.-K. & Jakkula, E. Estimation of carrier frequencies of autosomal and X-linked recessive genetic conditions based on gnomAD v4.0 data in different ancestries. *Genet. Med.* **27**, 101304 (2025).
15. Committee Opinion No. 690: Carrier Screening in the Age of Genomic Medicine. *Obstet. Gynecol.* **129**, e35–e40 (2017).

16. Chokoshvili, D., Vears, D. & Borry, P. Expanded carrier screening for monogenic disorders: where are we now? *Prenat. Diagn.* **38**, 59–66 (2018).
17. Saier, C. *et al.* TREAT: systematic and inclusive selection process of genes for genomic newborn screening as part of the Screen4Care project. *Orphanet J. Rare Dis.* **20**, 231 (2025).
18. Lazarin, G. A. *et al.* Systematic classification of disease severity for evaluation of expanded carrier screening panels. *PLoS One* **9**, e114391 (2014).
19. Arjunan, A. *et al.* Evaluation and classification of severity for 176 genes on an expanded carrier screening panel. *Prenat. Diagn.* **40**, 1246–1257 (2020).
20. Varni, J. W., Seid, M. & Kurtin, P. S. PedsQL™ 4.0: Reliability and Validity of the Pediatric Quality of Life Inventory™ Version 4.0 Generic Core Scales in Healthy and Patient Populations. *Med. Care* **39**, 800–812 (2001).
21. Varni, J. W. *et al.* The PedsQL™ Infant Scales: feasibility, internal consistency reliability, and validity in healthy and ill infants. *Qual. Life Res.* **20**, 45–55 (2011).
22. Gargano, M. A. *et al.* The Human Phenotype Ontology in 2024: phenotypes around the world. *Nucleic Acids Res.* **52**, D1333–D1346 (2024).
23. Murphy, K. B., Schilder, B. M. & Skene, N. G. Using GPT-4 to annotate the severity of all phenotypic abnormalities within the human phenotype ontology. *Front. Digit. Health* **8**, 1794934 (2026).
24. Wang, T., Scuffham, P., Byrnes, J., Delatycki, M. B. & Downes, M. An overview of reproductive carrier screening panels for autosomal recessive and/or X-linked conditions: How much do we know? *Prenat. Diagn.* **43**, 1416–1424 (2023).
25. Guha, S. *et al.* Laboratory testing for preconception/prenatal carrier screening: A technical standard of the American College of Medical Genetics and Genomics (ACMG). *Genet. Med.* **26**, 101137 (2024).
26. Gruzin, M. J. *et al.* Optimizing gene panels for equitable reproductive carrier screening: The Goldilocks approach. *Genet. Med.* **27**, 101387 (2025).
27. Kirk, E. P. *et al.* Gene selection for the Australian reproductive genetic carrier screening project ("Mackenzie's Mission"). *Eur. J. Hum. Genet.* **29**, 79–87 (2021).
28. Kirk, E. P. *et al.* Nationwide, Couple-Based Genetic Carrier Screening. *N. Engl. J. Med.* **391**, 1877–1889 (2024).
29. Minten, T. *et al.* Data-driven consideration of genetic disorders for global genomic newborn screening programs. *Genet. Med.* **27**, 101443 (2025).
30. Sajko, M. C. *et al.* Towards a patient-centred classification of genetic disease severity. *Eur. J. Hum. Genet.* **33**, 1520–1529 (2025).
31. de Koning, M. A. *et al.* Experiences of Dutch parents undergoing prenatal genomic testing for fetal structural anomalies: A prospective qualitative analysis. *J. Genet. Couns.* **35**, e70187 (2026).
32. Wu, K. *et al.* An automated framework for assessing how well LLMs cite relevant medical references. *Nat. Commun.* **16**, 3615 (2025).
33. Cacheiro, P. *et al.* Lethal phenotypes in Mendelian disorders. *Genet. Med.* **26**, 101141 (2024).
34. Rapaport, F. *et al.* Negative selection on human genes underlying inborn errors depends on disease outcome and both the mode and mechanism of inheritance. *Proc. Natl. Acad. Sci. U. S. A.* **118**, e2001248118 (2021).
35. Ghiossi, C. E., Goldberg, J. D., Haque, I. S., Lazarin, G. A. & Wong, K. K. Clinical Utility of Expanded Carrier Screening: Reproductive Behaviors of At-Risk Couples. *J. Genet. Couns.* **27**, 616–625 (2018).
36. De Vries, I. A. C. *et al.* Prevalence of feeding disorders in children with cleft palate only: a retrospective study. *Clin. Oral Investig.* **18**, 1507–1515 (2014).
37. Hocevar-Boltezar, I., Jarc, A. & Kozelj, V. Ear, nose and voice problems in children with orofacial clefts. *J. Laryngol. Otol.* **120**, 276–281 (2006).
38. Gorodkin, J. Comparing two K-category assignments by a K-category correlation coefficient. *Comput. Biol. Chem.* **28**, 367–374 (2004).
39. Vasilevsky, N. A. *et al.* Mondo: integrating disease terminology across communities. *Genetics* **232**, iyaf215 (2026).

40. DiStefano, M. T. *et al.* The Gene Curation Coalition: A global effort to harmonize gene–disease evidence resources. *Genet. Med.* **24**, 1732–1742 (2022).
41. Miller, D. T. *et al.* ACMG SF v3.2 list for reporting of secondary findings in clinical exome and genome sequencing: A policy statement of the American College of Medical Genetics and Genomics (ACMG). *Genet. Med.* **25**, 100866 (2023).
42. Yatsenko, S. A. *et al.* The Human Intolerome: A curated database to prioritize genomic variants in stillbirth, pregnancy loss, and neonatal death. *Genet. Med.* **28**, 102546 (2026).
43. Seal, R. L. *et al.* Genenames.org: the HGNC and PGNC resources in 2026. *Nucleic Acids Res.* **54**, D1098–D1107 (2026).

**Table 1.** Quality of Life (PedsQL) Domain Distribution and Impact Across Severity Tiers. N Phenotypes is the number of phenotypes in each tier; percentages are shares of all QoL domain mappings among the QoL-affected phenotypes of the tier, and a phenotype can map to several domains. For Tier 4 the percentages come from its 44 QoL-affected phenotypes, all infertility terms; the other 1,473 Tier 4 phenotypes have no affected domain.

| Tier | N Phenotypes | Physical Functioning % | School/Cognitive Functioning % | Physical Symptoms % | Emotional Functioning % | Social Functioning % |
|---|---|---|---|---|---|---|
| Tier 1 | 272 | 47.7 | 31.0 | 13.0 | 4.2 | 4.2 |
| Tier 2 | 2,966 | 71.9 | 9.3 | 15.1 | 1.8 | 1.8 |
| Tier 3 | 2,351 | 49.8 | 19.3 | 9.2 | 10.2 | 11.5 |
| Tier 4 | 1,517 | 30.9 | 2.5 | 4.9 | 38.3 | 23.5 |

**Table 2.** Severity of gene–disease pairs by HPO body system. Body systems are the 23 children of Phenotypic abnormality (HP:0000118) in HPO release 2026-09-01. Severity was assigned by the hierarchical rules applied to the tier counts of each pair (Methods, Equation 7). A pair is counted under every body system that contains at least one of its phenotypes kept by the frequency filter (phenotypes with no frequency annotation, or with a highest annotated frequency of at least 30%), so a pair can appear in several rows. Analysis of 6,709 classified pairs; pairs supported only by a Limited GenCC classification are excluded. Rows are sorted by Profound share and each row sums to 100%.

| Body System (HPO Parent Term) | N pairs | Profound % | Severe % | Moderate % | Mild % |
|---|---|---|---|---|---|
| Abnormality of prenatal development or birth | 620 | 69.5 | 25.3 | 3.5 | 1.6 |
| Growth abnormality | 2,223 | 57.2 | 38.3 | 4.3 | 0.2 |
| Abnormality of the nervous system | 4,247 | 55.1 | 37.0 | 7.4 | 0.5 |
| Abnormality of the breast | 215 | 54.9 | 35.3 | 9.3 | 0.5 |
| Abnormality of head or neck | 3,294 | 54.7 | 35.5 | 8.0 | 1.8 |
| Abnormality of the ear | 1,750 | 53.3 | 35.0 | 11.5 | 0.2 |
| Abnormality of the eye | 2,906 | 53.2 | 35.1 | 11.7 | 0.1 |
| Abnormality of the digestive system | 2,517 | 52.9 | 38.1 | 8.0 | 1.0 |
| Abnormality of the musculoskeletal system | 4,249 | 52.0 | 41.0 | 6.7 | 0.3 |
| Abnormality of the respiratory system | 1,522 | 50.1 | 41.5 | 8.2 | 0.3 |
| Abnormality of the voice | 259 | 49.4 | 48.3 | 2.3 | 0.0 |
| Abnormality of limbs | 2,273 | 48.4 | 45.8 | 5.4 | 0.4 |
| Abnormality of the genitourinary system | 2,165 | 46.1 | 36.8 | 10.6 | 6.6 |
| Abnormal cellular phenotype | 641 | 42.7 | 44.9 | 11.4 | 0.9 |
| Abnormality of the integument | 2,310 | 42.0 | 38.1 | 16.4 | 3.5 |
| Abnormality of the endocrine system | 809 | 41.4 | 40.4 | 13.3 | 4.8 |
| Abnormality of the cardiovascular system | 2,269 | 41.1 | 44.6 | 12.0 | 2.3 |
| Abnormality of metabolism / homeostasis | 2,379 | 40.8 | 41.0 | 14.1 | 4.1 |
| Abnormality of the immune system | 1,583 | 38.9 | 42.5 | 17.7 | 0.9 |
| Constitutional symptom | 719 | 30.5 | 56.5 | 11.7 | 1.4 |
| Abnormality of blood and blood-forming tissues | 1,226 | 29.4 | 43.1 | 19.7 | 7.9 |
| Neoplasm | 502 | 23.5 | 44.2 | 31.5 | 0.8 |
| Abnormality of the thoracic cavity | 14 | 21.4 | 78.6 | 0.0 | 0.0 |
| Overall (all 6,709 classified gene–disease pairs) | 6,709 | 36.3 | 37.5 | 21.4 | 4.8 |

## Figure legends

**Figure 1. AI-based Workflow for Automated Phenotype and Gene Severity Classification**. Schematic overview of the complete classification pipeline demonstrating: (1) Phenotype classification criteria integrating ACMG severity tiers and ACOG criteria including PedsQL domains, age of onset, and treatment availability; (2) ReAct agent framework with iterative thought-action-observation loops executing five phases: contextualization, query generation, evidence retrieval, adaptive expansion and classification; (3) Source verification of the cited claims; and (4) Gene-level severity aggregation algorithm producing final classifications of Profound, Severe, Moderate, or Mild based on associated phenotype profiles.

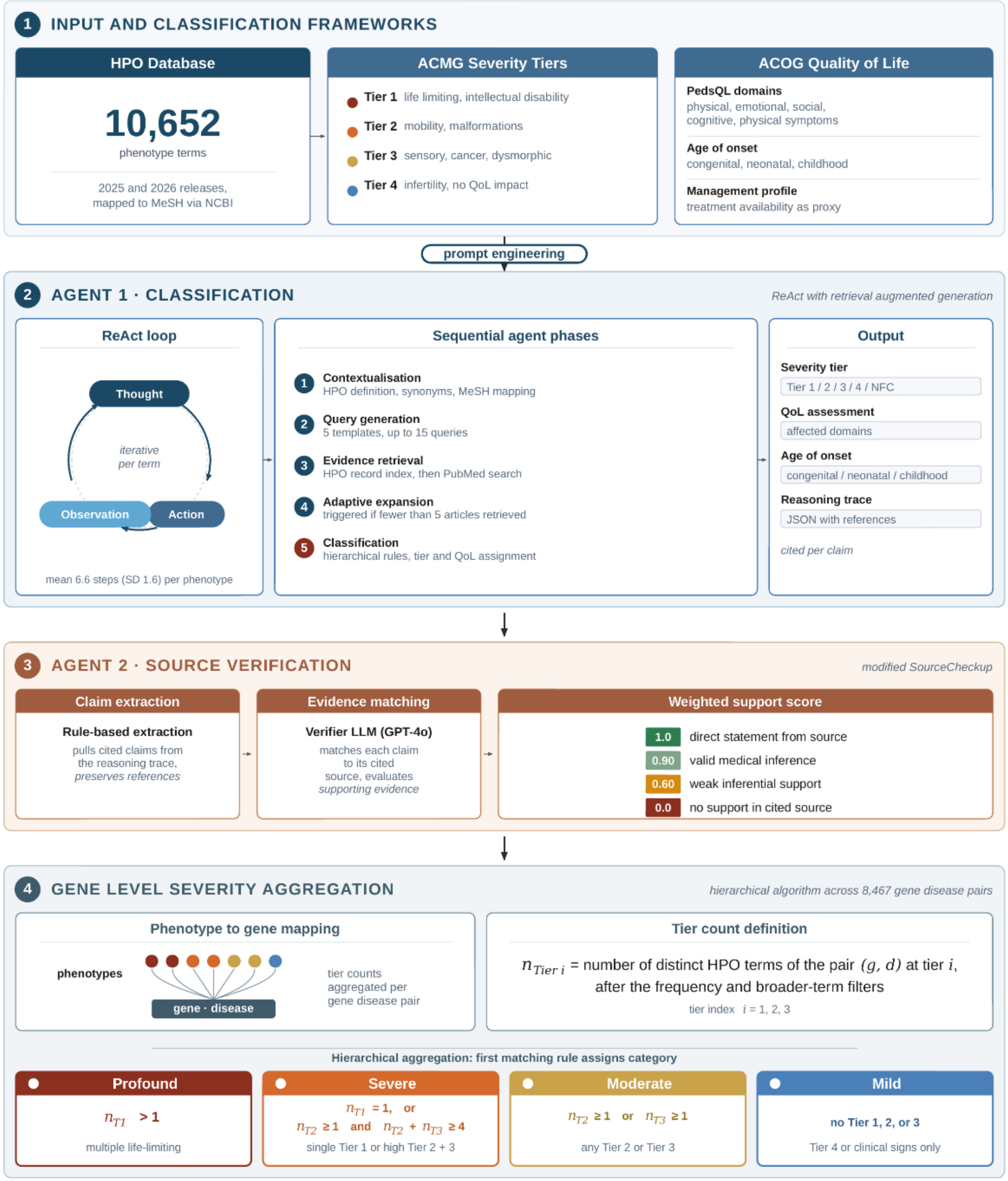

**Figure 2. HPO Classification Model Performance Metrics**. Agent output for the 898 scored expert-labelled terms. (a) Overall performance: accuracy, weighted and macro-averaged precision, recall and F1, and Matthews correlation coefficient. (b) Precision, recall and F1 for each of the 16 categories (13 tiered categories, Clinical Sign or Laboratory Abnormality, Out of Scope and NFC). (c) Predicted against true severity tier (jittered); Clinical Sign or Laboratory Abnormality, Out of Scope and NFC terms form the non-tiered class (labelled NFC in the panel). (d) Precision, recall and F1 for each tier and the non-tiered class. (e) Agent against the no-retrieval baseline (same model and prompt, no PubMed evidence) on the same terms: category accuracy and kappa, tier accuracy and quadratic-weighted kappa; error bars are 95% bootstrap confidence intervals (2,000 resamples). (f) Category accuracy of the agent and the baseline in each category; the number of expert-labelled terms is given in brackets.

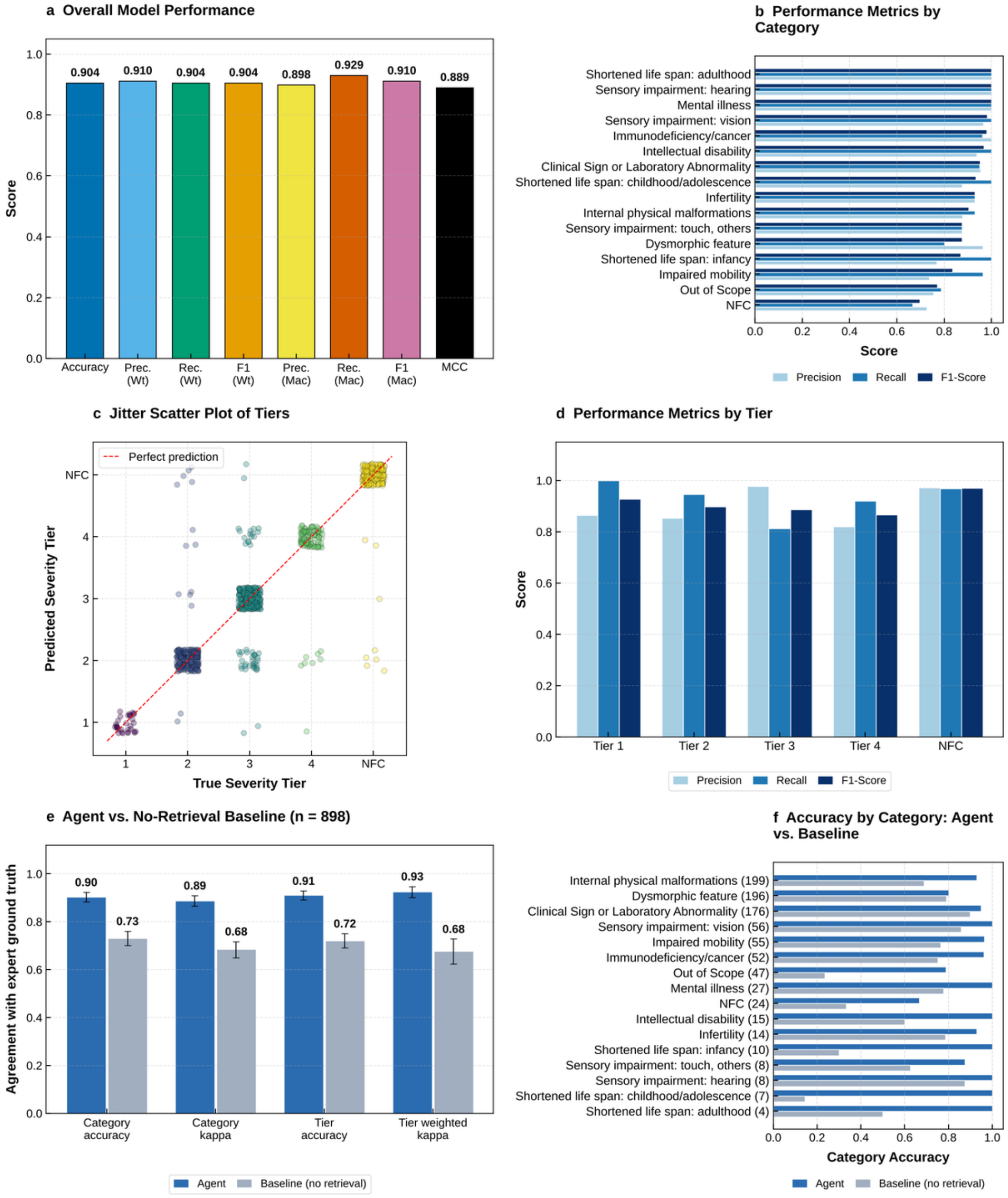

**Figure 3. Severity Tiers, Categories, Management and Early Onset of the Classified HPO Terms.** (a) Distribution of severity tiers within each of the 16 categories (13 tiered categories and the three non-tiered designations); N is the number of phenotypes. (b) Prevalence of early onset designation in the 13 tiered categories; the panel title gives the chi-square statistic and P value. (c) Management profiles showing the distribution of intervention types in the 13 tiered categories. (d) Management profile across severity tiers: distribution of intervention types in each tier.

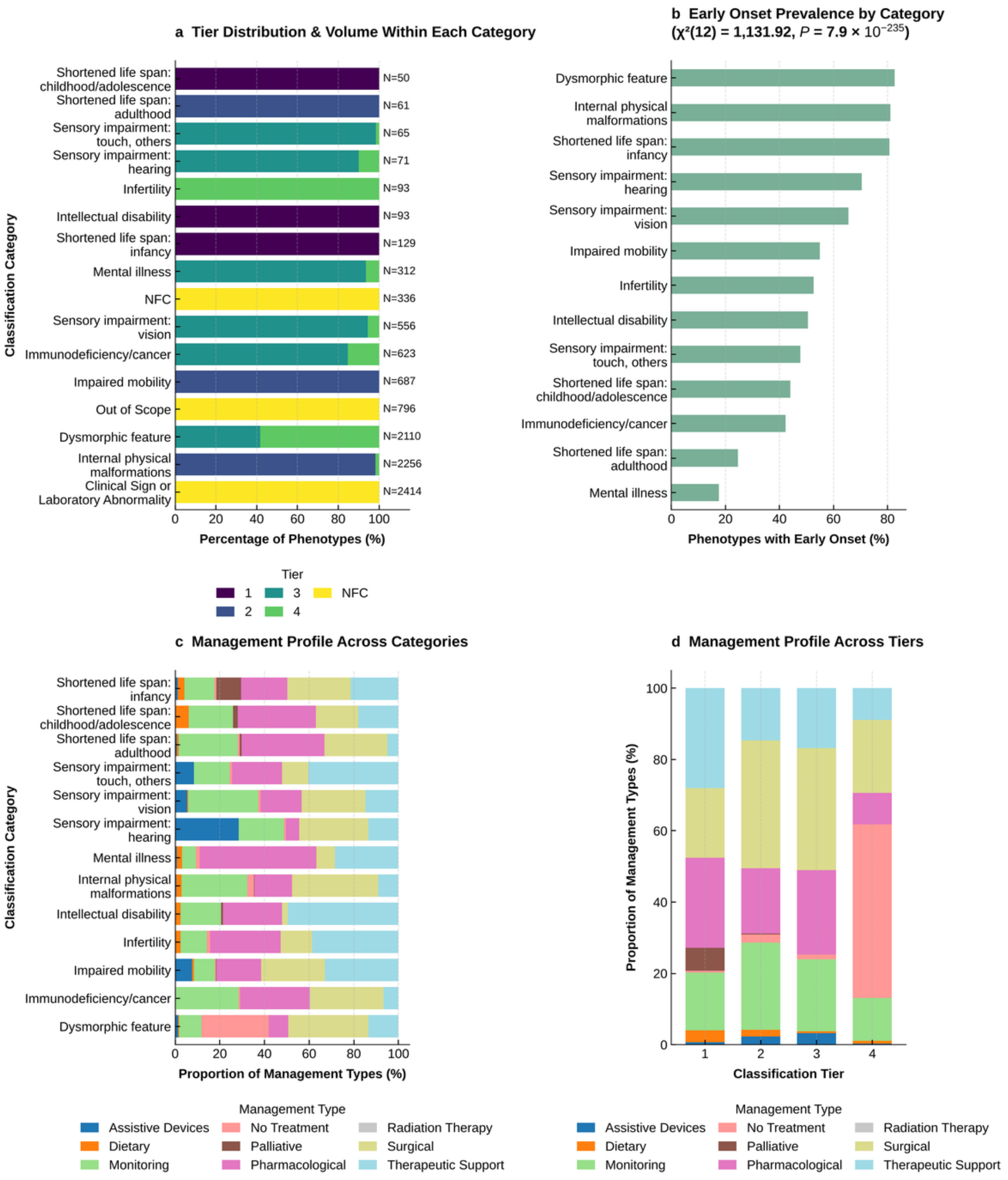

**Figure 4**. **Evidence Quality and Classification Confidence Metrics**. (a) Share of claims that were direct evidence, valid inferences, weak inferences (unlabelled segment) or unsupported, by severity tier. (b) The same shares in each of the 13 tiered categories. (c) Distribution of statement support scores by severity tier (violin plots; dashed lines mark the quartiles). (d) Distribution of statement support scores in each of the 13 tiered categories. Panel titles give the Kruskal–Wallis statistics and P values.

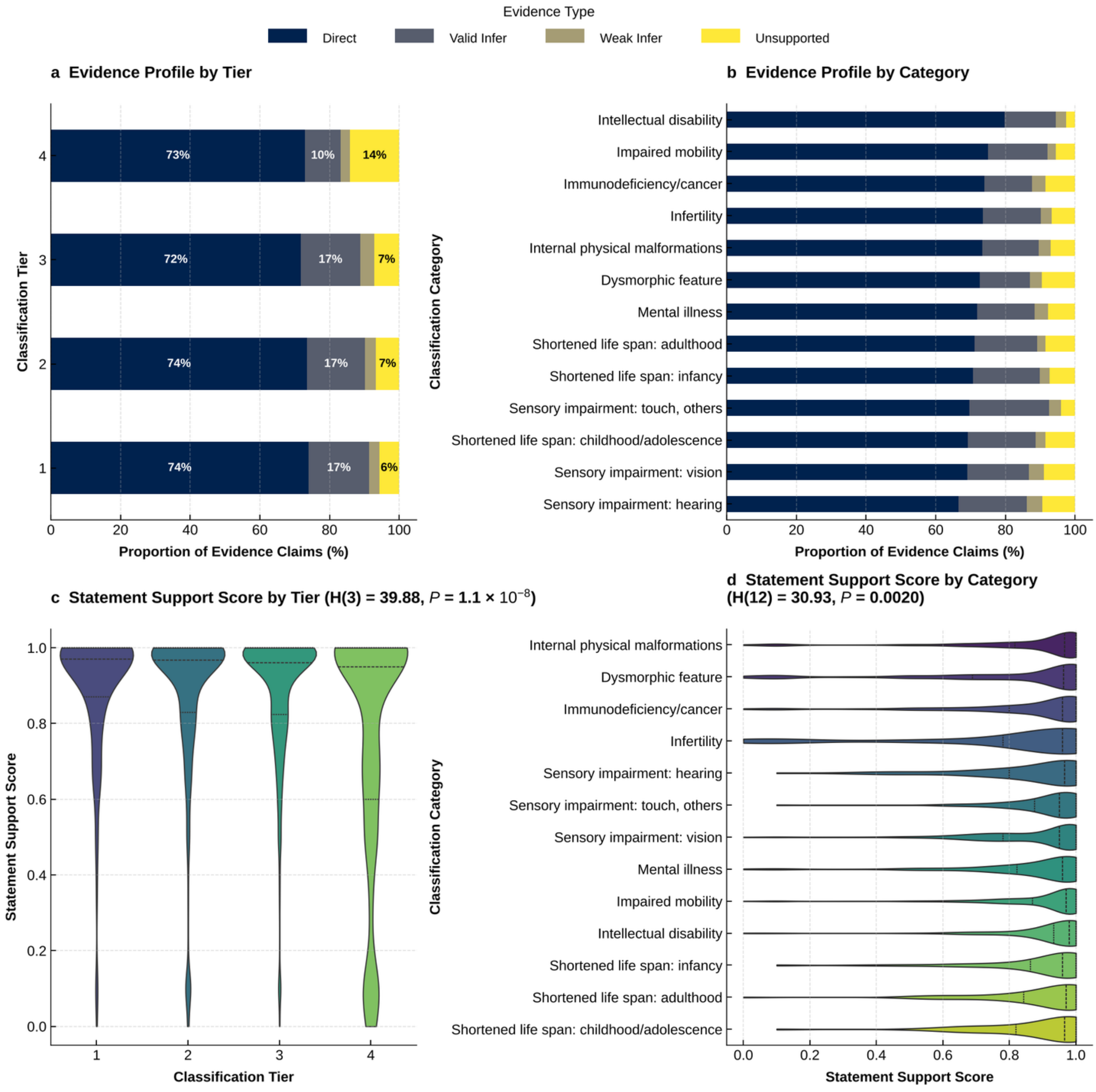

**Figure 5. Association of HPO Classifications with Quality-of-Life (QoL) Domain Impact Profiles**. (a) Share of PedsQL™ quality-of-life domain mappings in each severity tier; a phenotype can map to several domains, and all QoL-affected Tier 4 phenotypes are infertility terms (n = 44). (b) Share of quality-of-life domain mappings in each of the 13 tiered categories. Panel titles give the chi-square statistics and P values.

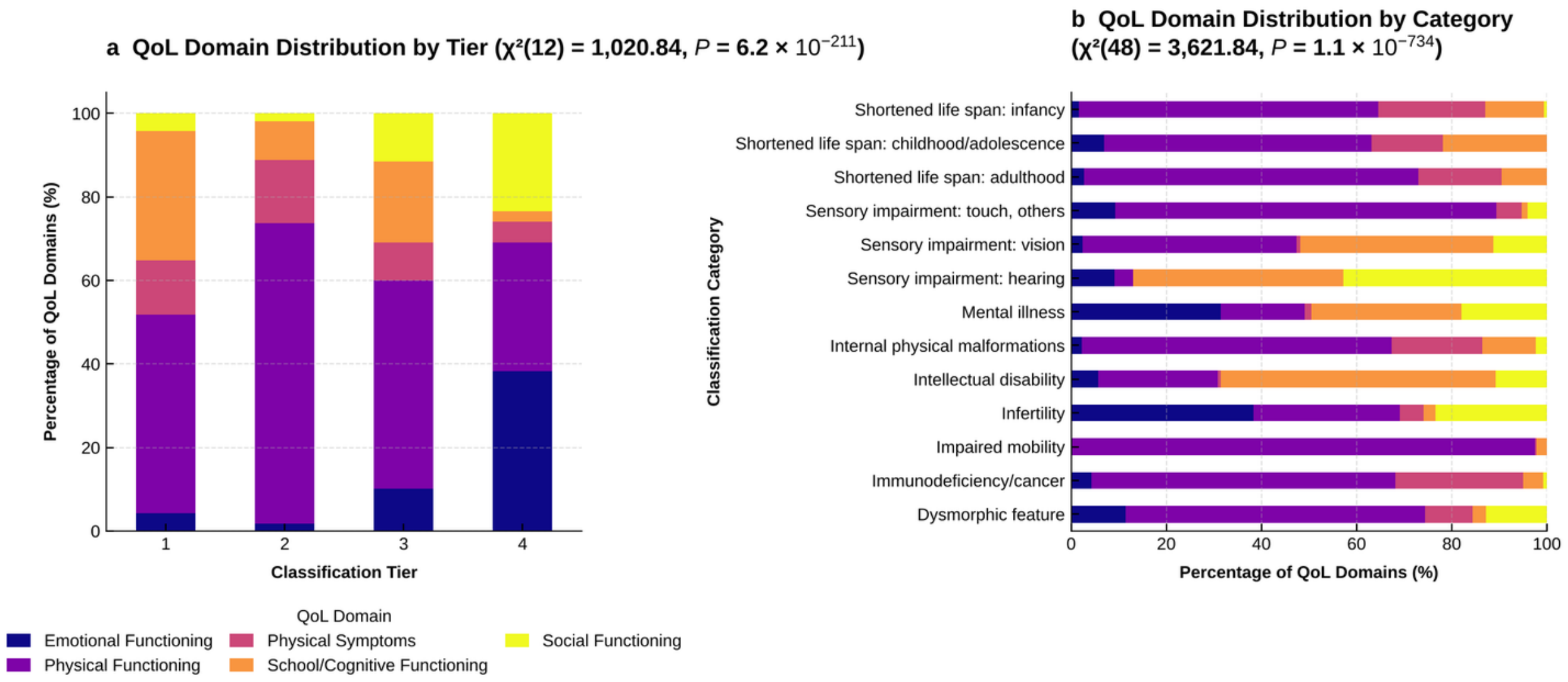

**Figure 6. Management Complexity and Quality of Life Domain Analysis of Predicted Severity Tiers**. (a) Correlation between QoL domain involvement and therapeutic intensity, with the fitted linear relationship and its 95% bootstrap confidence band overlaid on the data. (b) Overall frequency of PedsQL™ domains across all 10,652 classified phenotypes; a phenotype can affect several domains. (c) Tier-stratified management complexity showing the mean number of interventions for Tiers 1 to 4, with standard deviation error bars. (d) Category-specific management complexity with 95% confidence intervals (1.96 × standard error of the mean) across the 13 tiered categories. No treatment is not counted as an intervention. Panel titles give the regression and Kruskal–Wallis statistics and P values.

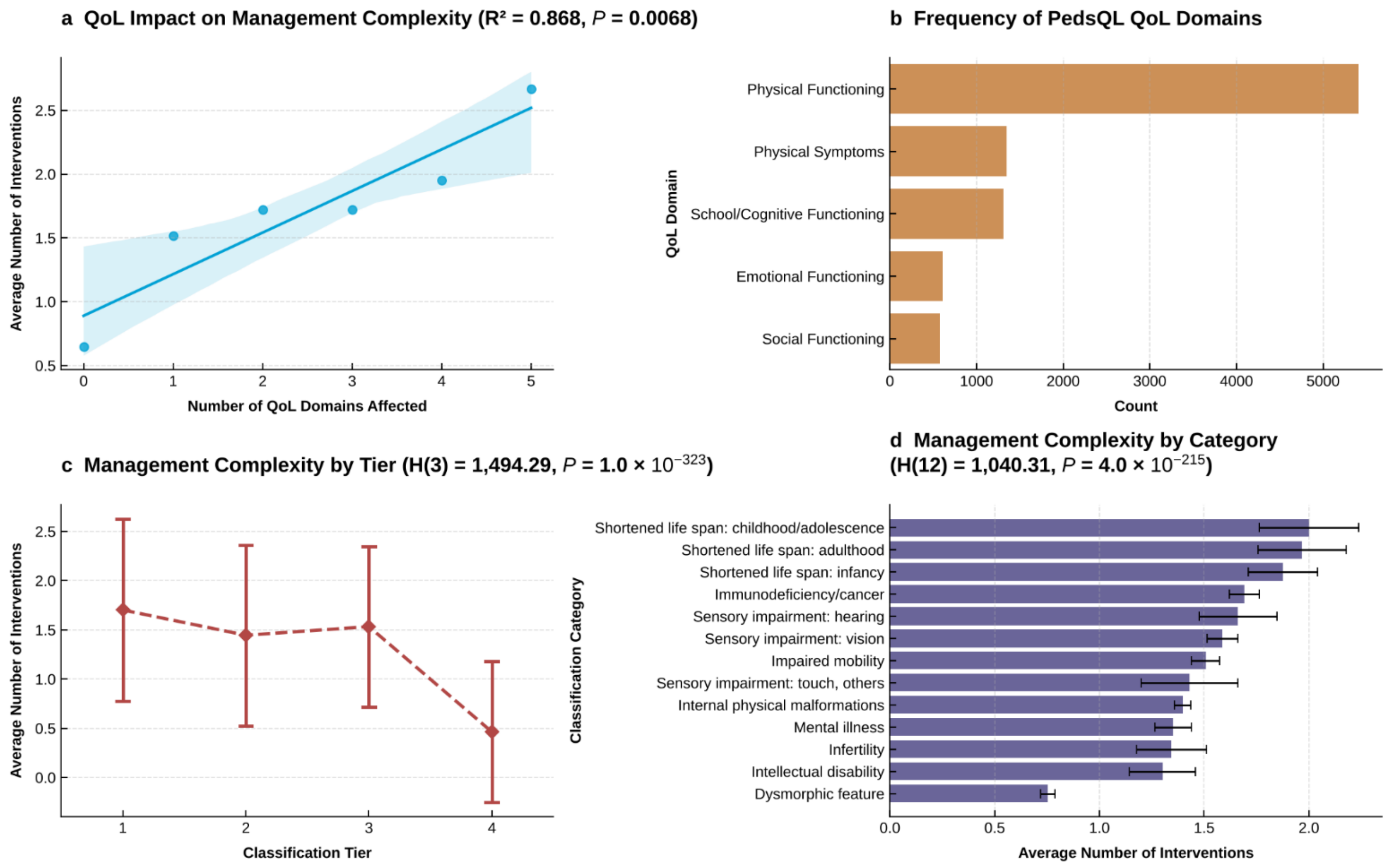

**Figure 7. Gene-Level Severity Classification and Validation.** (a) Severity class of 6,709 gene–disease pairs; pairs supported only by a Limited GenCC classification are excluded. (b) Severity across the 23 HPO body systems, ranked by Profound share; a pair is counted under every body system that holds one of its phenotypes. (c) Severity by mode of inheritance; complex means more than one mode or other inheritance terms only; 38 pairs without an inheritance annotation are not shown. (d) Severity of genes in four clinical gene lists against all 4,531 genes (base rate). List genes take the highest class among their listed conditions, and base-rate genes the highest over all their pairs; percentages at right are the Severe or Profound share. (e) Share of the same genes with a phenotype tiered as shortened life span in infancy or childhood/adolescence. n is the number of pairs (b, c) or of genes with a matched, classified condition (d). Numbers inside bars are percentages.

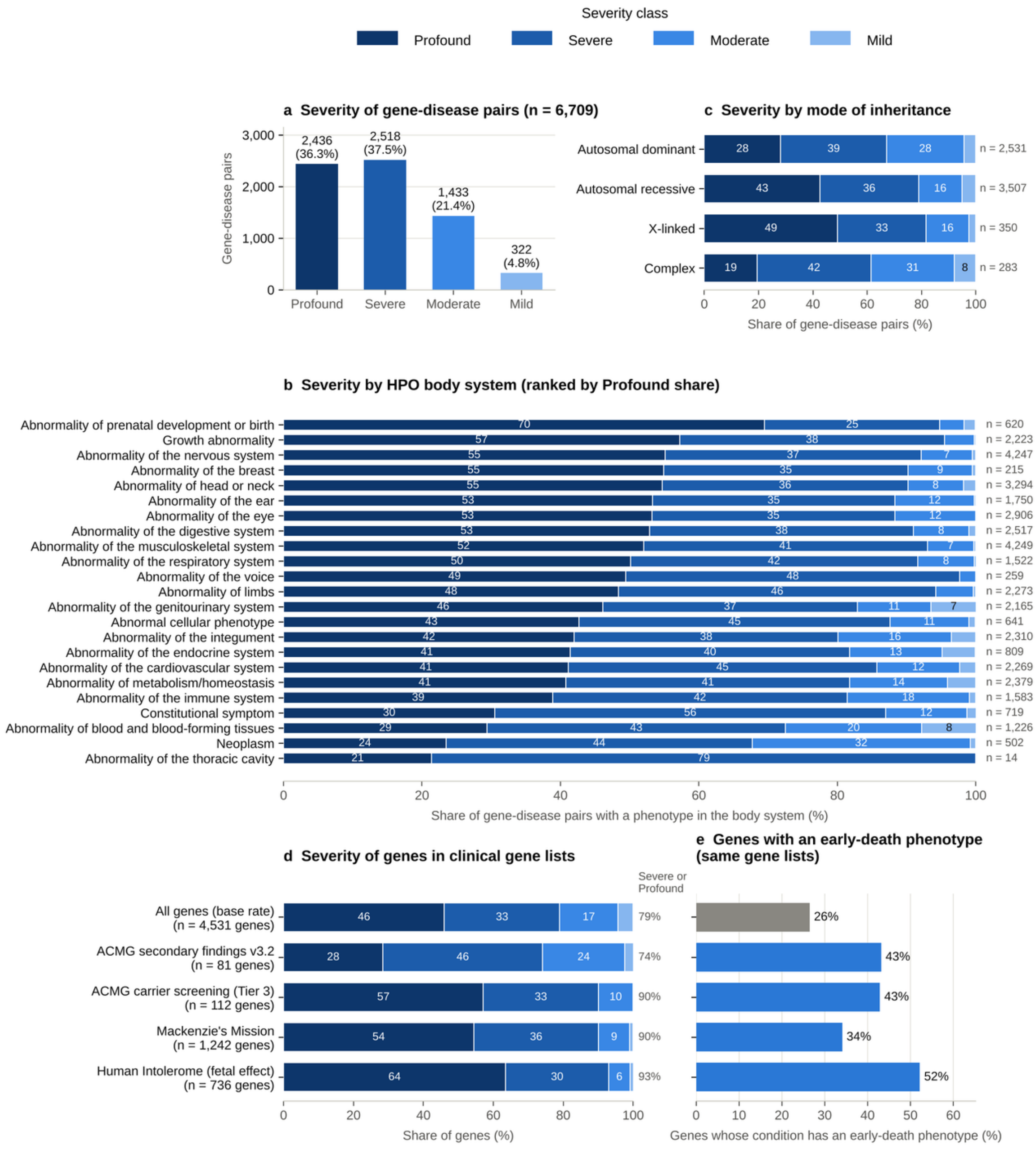

## Supplementary Information

# Large Language Model Agents for Evidence Based Genetic Disease Severity Classification

Tohid Ghasemnejad[1], Ahmadreza Argha[2], Mark Grosser[3], John Wang[4], Min Yang[5], Thantrira Porntaveetus[6], Tony Roscioli[7,8], Nigel H. Lovell[2], Mahmoud Aarabi*[9,10], Hamid Alinejad-Rokny*[1]

[1] UNSW BioMedical Machine Learning Lab (BML), School of Biomedical Engineering, UNSW Sydney, Sydney, NSW 2052, Australia

[2] School of Biomedical Engineering, UNSW Sydney, NSW 2052, Australia

[3] 23Strands, Pyrmont, Australia

[4] Division of Genetic and Genomic Medicine, Department of Pediatrics, University of Pittsburgh School of Medicine, Pittsburgh, PA, USA

[5] Shenzhen Institute of Advanced Technology, Shenzhen, China

[6] Center of Excellence in Precision Medicine and Digital Health, Department of Physiology, Faculty of Dentistry, Chulalongkorn University, Bangkok, Thailand

[7] New South Wales Health Pathology Genomics, Prince of Wales Hospital, Randwick, NSW, Australia

[8] Neuroscience Research Australia (NeuRA), University of New South Wales Sydney, NSW, Australia

[9] Medical Genetics and Genomics Laboratories, University of Pittsburgh Medical Center, Pittsburgh, PA, USA

[10] Departments of Pathology, and Obstetrics, Gynecology, and Reproductive Sciences, University of Pittsburgh School of Medicine, Pittsburgh, PA, USA

Correspondence to: H.A.R (h.alinejad@unsw.edu.au) and M.A. (mahmoud.aarabi@pitt.edu)

## Supplementary Tables

**Supplementary Table 1. Category Definitions for Phenotype Classification Framework.** The agent checks the categories in a fixed order and assigns the first that fits. Clinical Sign or Laboratory Abnormality, Out of Scope and NFC are treated as non-tiered in all analyses.

| Category | Tiers | Definition | Ref |
|---|---|---|---|
| Shortened life span: infancy | Tier 1 | A phenotype or condition that typically results in death or a significantly reduced life expectancy during the first two years of life (from birth to 24 months of age), including any condition in which death is likely before two years of age. | 1 |
| Shortened life span: childhood/adolescence | Tier 1 | A phenotype or condition that typically leads to death or a markedly reduced life expectancy during childhood (from 2 to 12 years) or adolescence (ages 13–18 years). | 2,3 |
| Shortened life span: adulthood | Tier 2 | A phenotype or condition that usually causes death or substantially shortens life expectancy during adulthood (after age 18 years). | 4,5 |
| Intellectual disability | Tier 1 | A phenotype or condition where there are significant limitations in intellectual functioning (such as problem-solving, reasoning, or learning) and in adaptive behaviors (practical, social, and conceptual skills), with onset before the age of 18. | 6–9 |
| Mental illness | Tier 3 | Mental illness characterized by clinically significant disturbance in an individual's cognition, emotional regulation, or behavior that reflects dysfunction in psychological, biological, or developmental processes underlying mental and behavioral functioning. These disturbances are associated with significant distress or disability in social, occupational, or other important areas of functioning. | 10–12 |
| Sensory impairment: vision | Tier 3 | Partial or complete loss of visual ability, including conditions that lead to reduced visual acuity (sharpness), field, or even total blindness. | 13,14 |
| Sensory impairment: hearing | Tier 3 | Partial or complete loss of the ability to hear, ranging from mild hearing loss to total deafness in one or both ears. | 15,16 |
| Internal physical malformations | Tier 2 | A congenital (present at or before birth) structural anomaly that affects the shape, formation, or location of an internal organ or tissue (such as the heart, kidneys, liver, lungs, or internal brain structures). These malformations are not visible externally and are typically detected by imaging, surgery, or autopsy. They often impact the function of the involved organ. | 17–19 |
| Impaired mobility | Tier 2 | Any phenotype where physical movement or locomotion is significantly limited, ranging from difficulty walking to complete inability to walk, run, or perform other motor activities, whether due to weakness, paralysis, stiffness, or other causes. | 20,21 |
| Sensory impairment: touch, others | Tier 3 | Impairment in the ability to perceive touch, pain, temperature, position sense (proprioception), or any other non-visual/non-auditory sensory functions, and deficits in smell or taste. | 22 |
| Immunodeficiency/cancer | Tier 3 | Immunodeficiency: A primary or secondary deficiency of immune function, resulting in increased risk or frequency of infections.<br>Cancer: Presence, risk, or history of malignant neoplastic conditions (abnormal growth of cells that can invade or spread). | 23 |
| Dysmorphic feature | Tier 3 | An abnormality of the external body or face's shape, size, or proportions that is observable to the naked eye or through standard physical exam. Dysmorphic features are external and visible, not hidden inside the body. | 24,25 |
| Infertility | Tier 4 | A condition where a person is unable to conceive children naturally, either due to lack of functional gametes, reproductive organs, or other causes that directly prevent reproduction. | 26–28 |
| Clinical Sign or Laboratory Abnormality | Non-tiered | Objective, observable findings that represent transient symptoms, temporary clinical manifestations, or specific laboratory test results rather than persistent functional disabilities. | - |
| Out of Scope: Acquired/Multifactorial | Non-tiered | Phenotypes that are primarily acquired through environmental factors, lifestyle, or aging, or have complex multifactorial etiology with typical adult onset. | - |
| Not Further Classified (NFC) | Non-tiered | Terms that do not align with any of the established severity tier categories or classification criteria. | - |

**Supplementary Table 2.** Quality of Life (PedsQL) Domain Distribution and Impact Across Severity Tiers and Phenotypic Categories (7,106 phenotypes in the 13 tiered categories)**.**

| Phenotypic Category | Tier(s) | N Total | N QoL-Affected | % Affected | Predominant QoL Domain(s) |
|---|---|---|---|---|---|
| Internal physical malformations | Tiers 2, 4 | 2,256 | 2,218 | 98.3 | Physical Functioning |
| Impaired mobility | Tier 2 | 687 | 687 | 100.0 | Physical Functioning |
| Shortened life span: adulthood | Tier 2 | 61 | 61 | 100.0 | Physical Functioning, Physical Symptoms |
| Shortened life span: infancy | Tier 1 | 129 | 129 | 100.0 | Physical Functioning, Physical Symptoms |
| Shortened life span: childhood/adolescence | Tier 1 | 50 | 50 | 100.0 | Physical Functioning, School/Cognitive Functioning |
| Intellectual disability | Tier 1 | 93 | 93 | 100.0 | School/Cognitive Functioning |
| Sensory impairment: touch, others | Tiers 3–4 | 65 | 64 | 98.5 | Physical Functioning |
| Sensory impairment: vision | Tiers 3–4 | 556 | 524 | 94.2 | Physical Functioning, School/Cognitive Functioning |
| Mental illness | Tiers 3–4 | 312 | 292 | 93.6 | School/Cognitive Functioning, Emotional Functioning |
| Sensory impairment: hearing | Tiers 3–4 | 71 | 64 | 90.1 | School/Cognitive Functioning, Social Functioning |
| Immunodeficiency/cancer | Tiers 3–4 | 623 | 526 | 84.4 | Physical Functioning, Physical Symptoms |
| Infertility | Tier 4 | 93 | 44 | 47.3 | Emotional Functioning, Physical Functioning |
| Dysmorphic feature | Tiers 3–4 | 2,110 | 878 | 41.6 | Physical Functioning |

**Supplementary Table 3.** PedsQL™ Infant Scales items used to define the quality-of-life domains (provided as a separate Excel file, 4_Supplementary_Table_3_PedsQL_items.xlsx).

## Supplementary Figures

**Figure S1. Large Language Model Selection and Optimization.** Systematic comparison of classification accuracy across six large language models (GPT-5, Grok-4, Gemini-2.5-pro, Claude-3.5-sonnet, GPT-4o, and GPT-4o-mini) evaluated at three temperature settings (0.0, 0.2, 0.5) on 14 expert-labelled HPO terms spanning the severity tiers and phenotypic categories. The panel shows classification accuracy and average processing time for each model and temperature combination.

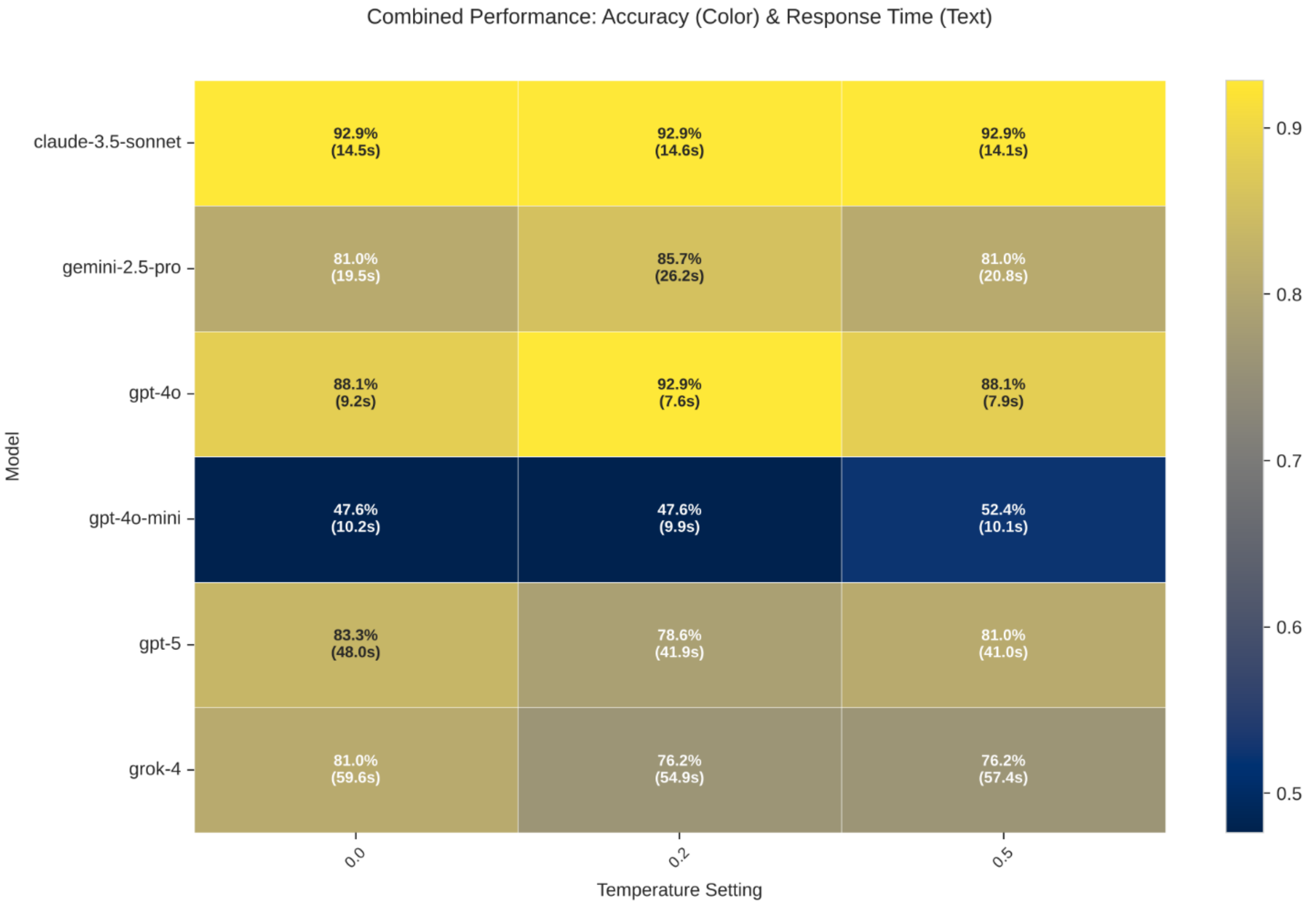

**Figure S2.** PedsQL Functional Impact Threshold and Associated Management Profiles Across Severity Tiers and Phenotypic Categories. **(a)** Category-specific quality of life status showing percentage of phenotypes meeting ACOG functional impact threshold versus unaffected phenotypes, stratified by phenotypic category. **(b)** Tier-stratified quality of life status showing the proportion of phenotypes in each tier meeting the QoL threshold. **(c)** Management profile comparison between QoL-affected and unaffected phenotypes, showing the distribution of intervention types (surgical, pharmacological, monitoring, therapeutic, dietary, assistive devices, palliative care, and no treatment) in each group. Panel titles give the chi-square statistics and P values.

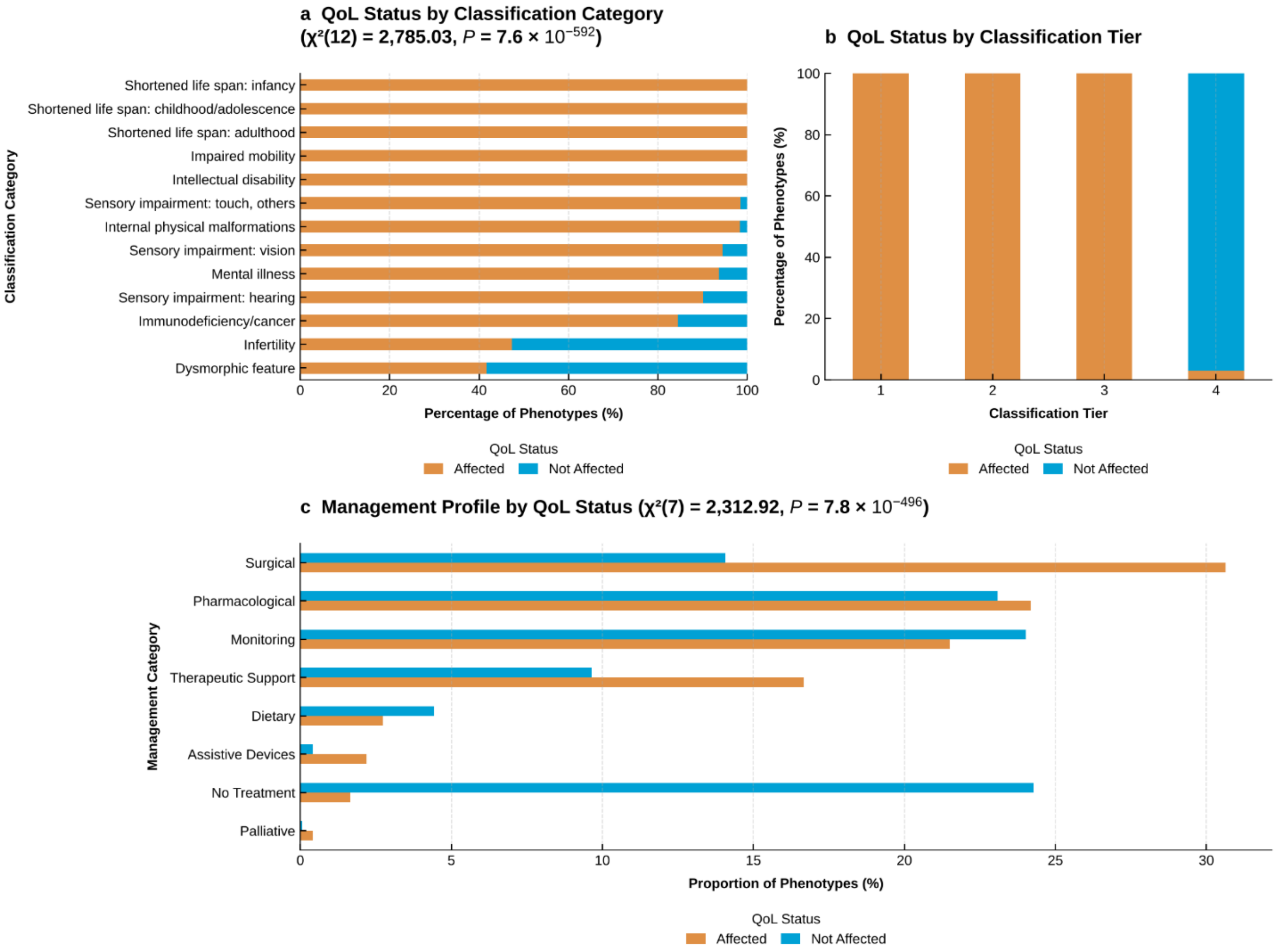

**Figure S3. Validation Sample, Category Errors, Evidence Support and Cost.** (a) Category distribution of the 898 expert-labelled terms (sample) against all classified terms (population), with the Population Stability Index (PSI = 0.0303). (b) Normalized confusion matrix of the agent on the 898 scored terms across 16 categories (rows: expert label; values are row percentages). (c) Distribution of statement support scores across the 10,651 scored classifications (dashed line: mean 0.82). (d) Cost per classification against the number of ReAct steps (Pearson r = 0.40, P = 2.5 × $10^{-410}$), with marginal distributions.

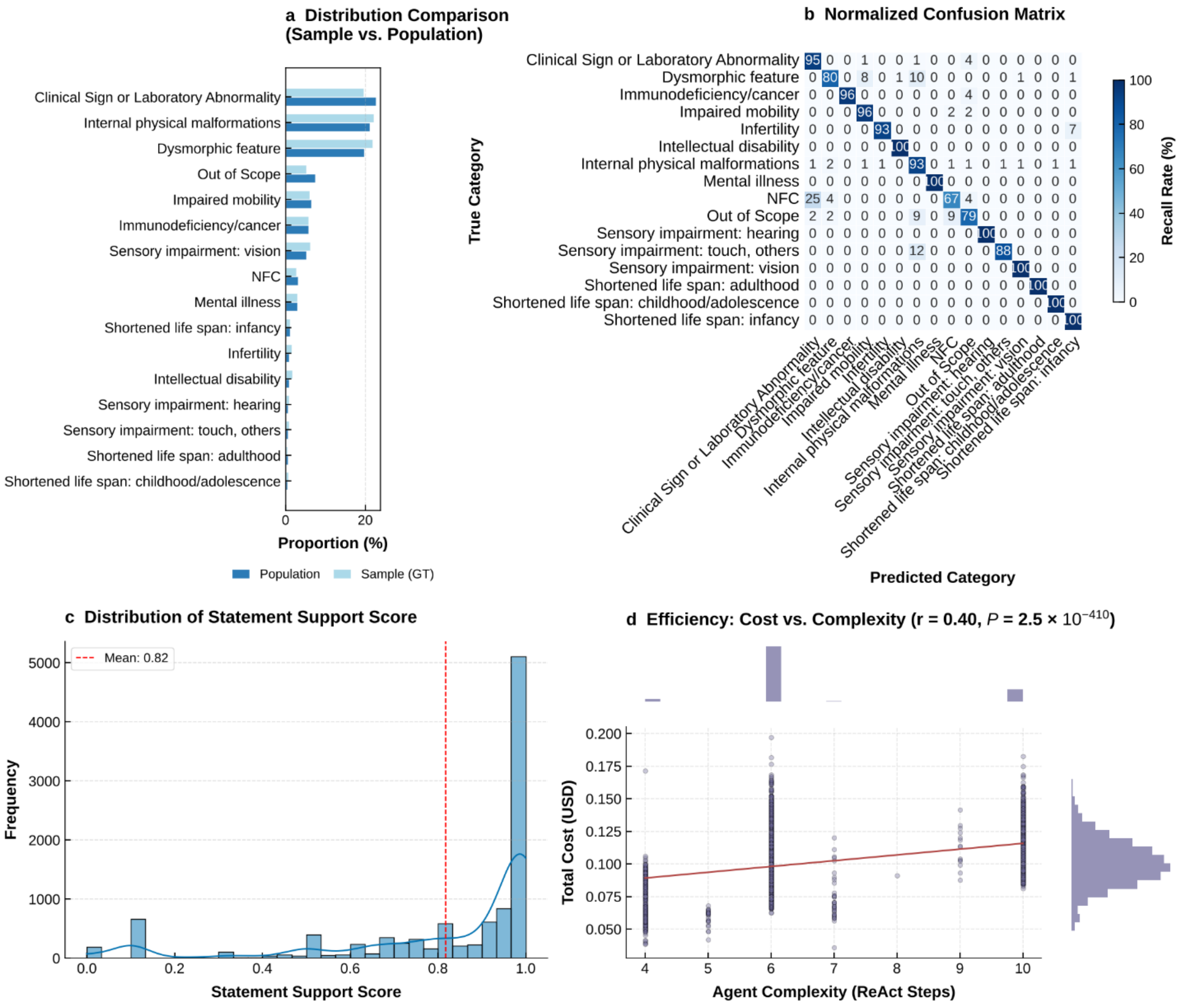

**Figure S4. Illustrative Example of Gene-Level Severity Classification.** Worked example demonstrating the complete classification pipeline for a representative gene–disease pair. The upper panel shows the gene summary including the gene and disease identifiers, inheritance pattern, GenCC classification and tier counts, with the associated phenotypes and their HPO identifiers, annotated frequencies, tier assignments, phenotypic categories and quality-of-life domains affected; the phenotype removed by the frequency filter is shown in grey. The middle panel presents the gene-level severity aggregation rule applied to the phenotype tier counts to derive the final classification. The lower panel shows the agent's saved reasoning for one phenotype: preliminary checks, quality-of-life functional impact, category, severity, management profile, final classification and conclusion, with the cited references and the result of source verification.

## PCDH15 and Usher syndrome type 1F

HGNC:14674 | MONDO:0011186 | OMIM:602083 | Autosomal recessive | GenCC: Definitive

T1: 0 T2: 2 T3: 2 T4: 0

### Associated phenotypes (5 annotated; 4 kept after the frequency filter)

| HPO ID | Phenotype | Frequency | Tier | Category | QoL domains affected |
|---|---|---|---|---|---|
| HP:0001751 | Abnormal vestibular function | 7/8 | 2 | Impaired mobility | Physical Functioning |
| HP:0031629 | Impaired tandem gait | 7/8 | 2 | Impaired mobility | Physical Functioning |
| HP:0000510 | Rod-cone dystrophy | 14/15 | 3 | Sensory Impairment – Vision | Physical, School/Cognitive, Social |
| HP:0008527 | Congenital sensorineural hearing impairment | 15/15 | 3 | Sensory Impairment – Hearing | School/Cognitive, Social |
| *HP:0001270* | *Motor delay (removed: below 30%)* | *1/8* | 2 | *Impaired mobility* | *Physical Functioning* |

**Pair severity: Severe.** Counted phenotypes: Tier 1 = 0, Tier 2 = 2, Tier 3 = 2. Rules, checked in order: Profound if Tier 1 > 1; Severe if Tier 1 = 1, or Tier 2 ≥ 1 and Tier 2 + Tier 3 ≥ 4; Moderate if Tier 2 ≥ 1 or Tier 3 ≥ 1; otherwise Mild. This pair meets the Severe rule (Tier 2 + Tier 3 = 4).

### Saved agent reasoning: Rod-cone dystrophy (HP:0000510)

Tier 3

#### Preliminary checks

**Origin type:** Inherited Genetic. **Phenotype type:** physiological function. **Evidence scope:** The evidence is about this phenotype specifically, with mentions of non-syndromic and syndromic forms but evaluated in isolation. **Body system:** visual system. **Onset:** adolescence or childhood. **Exclusion filters:** No exclusion filters apply; not a medical intervention, evidence shows inherited genetic causes.

#### QoL functional impact

**Affected domains:** Physical Functioning, School/Cognitive Functioning, Social Functioning.
**Evidence:** RP typically manifests with night blindness in adolescence, followed by concentric visual field loss, reflecting the principal dysfunction of rod photoreceptors; central vision loss occurs later in life due to cone dysfunction [Ref 2]. The most common form of RP is a rod-cone dystrophy, in which the first symptom is night blindness, followed by the progressive loss in the peripheral visual field in daylight, and eventually leading to blindness after several decades [Ref 3].
**Interpretation:** Vision loss from rod-cone dystrophy impairs daily physical activities such as navigation and self-care, affects school performance through reduced ability to read or learn visually, and limits social participation due to progressive blindness.

#### Category determination

**Category:** Sensory Impairment – Vision.
**Evidence:** Retinitis pigmentosa (RP) encompasses a group of inherited retinal dystrophies characterized by the primary degeneration of rod and cone photoreceptors [Ref 2].
**Rationale:** The phenotype involves partial or complete loss of visual ability through progressive degeneration leading to reduced visual field and eventual blindness, fitting sensory deficit in vision.

#### Severity assessment*

**Severity:** Severe.
**Evidence:** RP is a leading cause of visual disability, with a worldwide prevalence of 1:4000 [Ref 2]. Eventually leading to blindness after several decades [Ref 3].
**Multi-domain analysis:** Multiple domains are affected including physical functioning due to visual field loss impacting mobility and activities, school/cognitive due to central vision loss affecting learning, and social due to overall visual disability.

#### Management profile

**Categories:** Therapeutic Support, Monitoring/Surveillance.
**Summary:** Currently, there is no therapy that stops the evolution of the disease or restores the vision, so the visual prognosis is poor. The therapeutic approach is restricted to slowing down the degenerative process by sunlight protection and vitaminotherapy, treating the complications (cataract and macular edema), and helping patients to cope with the social and psychological impact of blindness [Ref 3]. We review management strategies for patients with RP, including counseling, visual rehabilitation, and current and emerging therapeutic options [Ref 2].

#### Final classification

**Tier:** 3. **Synthesis:** As a sensory impairment in vision with severe functional impact across multiple domains, it is classified in Tier 3.

#### Conclusion

[1] This condition is rod-cone dystrophy as supported by evidence stating 'An inherited retinal disease subtype in which the rod photoreceptors appear to be more severely affected than the cone photoreceptors' [Ref input_hpo].
[2] It is classified as Severe because 'eventually leading to blindness after several decades' [Ref 3].
[3] It belongs in the Sensory Impairment – Vision category due to 'the primary degeneration of rod and cone photoreceptors' [Ref 2].

#### References cited by the agent

[Ref input_hpo] Human Phenotype Ontology, HP:0000510 Rod-cone dystrophy, term definition.
[Ref 2] Verbakel SK, van Huet RAC, Boon CJF, et al. Non-syndromic retinitis pigmentosa. Prog Retin Eye Res. 2018;66:157–186. doi:10.1016/j.preteyeres.2018.03.005
[Ref 3] Hamel C. Retinitis pigmentosa. Orphanet J Rare Dis. 2006;1:40. doi:10.1186/1750-1172-1-40

**Source verification: 4 of 4 statements directly supported by the cited sources (support score 1.0).**

**Within each category, a phenotype rated as affecting quality of life takes the tier of its category, and a phenotype rated as not affecting it is placed in Tier 4. Frequency: highest annotated frequency of the phenotype in this disease (HPO release 2026-09-01); phenotypes below 30% are removed before the tiers are counted. The text of the lower panel is copied unchanged from the agent's saved output for this phenotype.*

## References

1. Newborn mortality. WHO https://www.who.int/news-room/fact-sheets/detail/newborn-mortality (accessed 13 October 2025).
2. Under-five mortality. UNICEF https://data.unicef.org/topic/child-survival/under-five-mortality/ (accessed 13 October 2025).
3. Child mortality (under 5 years). WHO https://www.who.int/news-room/fact-sheets/detail/child-mortality-under-5-years (accessed 13 October 2025).
4. What is the definition of adult mortality? World Bank https://datahelpdesk.worldbank.org/knowledgebase/articles/114956-what-is-the-definition-of-adult-mortality (accessed 13 October 2025).
5. Adult mortality rate (probability of dying between 15 and 60 years per 1000 population). WHO https://www.who.int/data/gho/indicator-metadata-registry/imr-details/64 (accessed 13 October 2025).
6. Disorders of intellectual development. ICD-11 https://icd.who.int/browse/2025-01/mms/en#605267007 (accessed 13 October 2025).
7. Papazoglou, A., Jacobson, L. A., McCabe, M., Kaufmann, W. & Zabel, T. A. To ID or Not to ID? Changes in Classification Rates of Intellectual Disability Using DSM-5. *Intellect. Dev. Disabil.* **52**, 165–174 (2014).
8. Maritska, Z., Prananjaya, B. A., Parisa, N. P., Suryani, P. R. & Hartati, M. D. An Insight into Genetic Conditions Associated with Intellectual Disability in Indonesian Population. *Int. J. Neuropsychopharmacol.* **28**, i186–i187 (2025).
9. Marrus, N. & Hall, L. Intellectual Disability and Language Disorder. *Child Adolesc. Psychiatr. Clin. N. Am.* **26**, 539–554 (2017).
10. Mental, behavioural or neurodevelopmental disorders. ICD-11 https://icd.who.int/browse/2025-01/mms/en#334423054 (accessed 13 October 2025).
11. Reed, G. M. What’s in a name? Mental disorders, mental health conditions and psychosocial disability. *World Psychiatry* **23**, 209–210 (2024).
12. Volpe, U. The development of the ICD-11 chapter on mental disorders. *S. Afr. J. Psychiatry* **23**, 1105 (2017).
13. Leat, S. J., Legge, G. E. & Bullimore, M. A. What Is Low Vision? A Re-evaluation of Definitions. *Optom. Vis. Sci.* **76**, 198–211 (1999).
14. Blindness and vision impairment. WHO https://www.who.int/news-room/fact-sheets/detail/blindness-and-visual-impairment (accessed 13 October 2025).
15. Davis, A. C. & Hoffman, H. J. Hearing loss: rising prevalence and impact. *Bull. World Health Organ.* **97**, 646–646A (2019).
16. Deafness and hearing loss. WHO https://www.who.int/news-room/fact-sheets/detail/deafness-and-hearing-loss (accessed 13 October 2025).
17. Pitt, D. B. A Study of Congenital Malformations: Part I. *Aust. N. Z. J. Obstet. Gynaecol.* **2**, 23–30 (1962).
18. Congenital disorders. WHO https://www.who.int/news-room/fact-sheets/detail/birth-defects (accessed 13 October 2025).
19. Corner, B. D. Congenital malformations. Clinical considerations. *Med. J. South-West* **77**, 46–52 (1962).
20. Habib, H. & Keller, E. J. Disability in Interventional Radiology. *Semin. Interv. Radiol.* **38**, 500–503 (2021).
21. Kittelson, T., Kittelson-Aldred, A., Justad, J. M., Hoffman, L. A. & Coombs, N. C. The Montana Postural Care Project: A pilot study implementing posture care management in a rural, low-resource region. *Heliyon* **10**, e31752 (2024).
22. Mak-Yuen, Y. Y. K., Matyas, T. A. & Carey, L. M. Characterizing Touch Discrimination Impairment from Pooled Stroke Samples Using the Tactile Discrimination Test: Updated Criteria for Interpretation and Brief Test Version for Use in Clinical Practice Settings. *Brain Sci.* **13**, 533 (2023).
23. Abolhassani, H., Wang, Y., Hammarström, L. & Pan-Hammarström, Q. Hallmarks of Cancers: Primary Antibody Deficiency Versus Other Inborn Errors of Immunity. *Front. Immunol.* **12**, 720025 (2021).
24. Claes, P. *et al.* Dysmorphometrics: the modelling of morphological abnormalities. *Theor. Biol. Med. Model.* **9**, 5 (2012).
25. Greene, C. L. *et al.* Clinical and biochemical footprints of inherited metabolic diseases: XVII. Dysmorphisms. *Mol. Genet. Metab.* **144**, 109001 (2025).
26. Infertility. WHO https://www.who.int/news-room/fact-sheets/detail/infertility (accessed 13 October 2025).
27. Zegers-Hochschild, F. *et al.* The International Glossary on Infertility and Fertility Care, 2017. *Hum. Reprod.* **32**, 1786–1801 (2017).
28. Hwang, K. *et al.* Mendelian genetics of male infertility. *Ann. N. Y. Acad. Sci.* **1214**, E1–E17 (2010).